\documentclass[11pt]{article}
\usepackage[english]{babel}
\input{epsf}
\usepackage{mathtext}
\usepackage{graphics,dcolumn,bm,float}
\usepackage{amssymb,amsmath,rotate,color}
\usepackage{times}
\newcommand{\ca}{{\cal A}}
\newcommand{\hhh}{{\cal H}}
\newcommand{\ii}{{\cal I}}
\newcommand{\CN}{{\cal N}}
\newcommand{\ee}{{\cal E}}
\newcommand{\be}{\begin{equation}}
\newcommand{\ene}{\end{equation}}
\newcommand{\ba}{\begin{array}}
\newcommand{\ea}{\end{array}}

\newcommand{\bsigma}{\mbox{\boldmath$\sigma$}}
\newcommand{\btau}{\mbox{\boldmath$\tau$}}

\begin{document}

\title{Transport properties of spin-triplet superconducting monolayer $MoS_2$}
\author{M. Khezerlou\footnote{m.khezerlou@urmia.ac.ir}, H. Goudarzi\footnote{Corresponding author; h.goudarzi@urmia.ac.ir, goudarzia@phys.msu.ru}\\
\footnotesize\textit{Department of Physics, Faculty of Science, Urmia University, Urmia, P.O.Box: 165, Iran}}
\date{}
\maketitle

\begin{abstract}
The quantum transport properties of graphene and monolayer $MoS_2$ superconductor heterostructures has been of considerable importance in the recent few years. Layered nature of molybdenum disulfide permits the superconducting correlation induction. Moreover, peculiar dynamical features of monolayer $MoS_2$, such as valence band spin-splitting in the nondegenerate $K$ and $K'$ valleys originated from strong spin-orbit coupling, and considerable direct band gap can make it potentially a useful material for electronics applications. Using the Dirac-like Hamiltonian of $MoS_2$ with taking into account the related mass asymmetry and topological contributions, we investigate the effect of spin-triplet $p$-wave pairing symmetry on the superconducting excitations, resulting in Andreev reflection process and Andreev bound state in the corresponding normal-superconductor (NS) and superconductor-normal-superconductor (SNS) structures, respectively. We study how the resulting subgap conductance and Josephson current are affected by the particular symmetry of order parameter. The signature of $p_x$-wave symmetry is found to decline the subgap superconducting energy excitations and, consequently, slightly suppress the Andreev reflection in the case of $p$-doped S region. The essential dynamical parameters $\lambda$ and $\beta$ of $MoS_2$ have significant effect on the both tunneling conductance and Josephson current. Particularly, the considered $p$-wave symmetry in the superconducting bound energies may feature the zero energy states at the interfaces. The critical current oscillations as a function of length of junction are obtained in the $p$-doped S region.
\end{abstract}

\section{INTRODUCTION}

The electron-like and hole-like quasiparticle excitations in the proximity superconducting two-dimensional (2D) materials, such as graphene $\cite{NG,CG}$ and monolayer molybdenum disulfide (ML-MDS) $\cite{ML,KS,RR}$ have triggered a massive interest in the related normal metal-superconductor and also Josephson junctions over the last few years. The role of various unconventional superconductor pairing symmetries in the transport properties of 2D structures (with graphene) consisting of relativistic charge carriers has been investigated by several authors (see, Refs. $\cite{LS,LB}$) in the recent years, following the premier work of Beenakker $\cite{B1,B2}$, which gave rise to discover specular Andreev reflection (AR) process. It has experimentally been shown that superconductivity may be induced by means of the proximity effect by placing a superconducting electrode near a graphene layer $\cite{HJ,DS,KD,MK,BB}$ or monolayer $MoS_2$ $\cite{G,TMS,Y,R}$. In addition to proximity inducing conventional $s$-wave pairing, in principle, the particular lattice symmetry of graphene and ML-MDS  (honeycomb with six Dirac points at the Brillouin zone) permits unconventional order parameters induction, such as spin-singlet $d$-wave or spin-triplet $p$-wave symmetries characterized by non-zero angular momentum of pair $l=2$ and $l=1$, respectively. The $p$-wave pairing has experimentally been observed in electronic system $Sr_2RuO_4$ indicating that its superconducting state has odd parity, breaks time-reversal symmetry and is spin-triplet $\cite{MM,NM,KSV,XM,L,MKN}$. Due to spontaneous time-reversal symmetry breaking in triplet pairing, the several anomalous transport phenomena, such as the anomalous Hall effect, polar Kerr effect for microwave radiation, anomalous Hall thermal conductivity, and anomalous photo and acousto-galvanic effects can be observed in the absence of external magnetic fields $\cite{NS,DM,SA}$. Very recently, some authors have also explored the chiral $p$-wave pairing in the various systems leading to anomalous transport phenomena and surface states $\cite{SA,BG}$. Moreover, it has been shown that the Majorana bands can be appeared in the vortex state of a chiral $p$-wave superconductor $\cite{J}$. Triplet pairing may also give rise to suppress the superconducting surface states in topological insulators $\cite{JL,GT,KM,BH,FK}$.

So far in the literature, unconventional $d$- and $p$-wave order parameter has much been studied in graphene, whereas a little attention has been paid to how unconventional pairing in monolayer $MoS_2$ influences the transport properties of related structures. Due to intrinsic massive Dirac gap (direct band gap of about $1.9\; eV$), strong spin-orbit coupling (SOC) caused by dichalcogenide heavy transition metal atom and resulting two nondegenerate $K$ and $K'$ valleys relative to spin-up and spin-down quasiparticles at the valence band $\cite{ZC,SZ,ZZ,CR,DL}$ (valley-contrasting spin-splitting $~0.1-0.5\; eV$ $\cite{ML}$) ML-MDS may exhibit dynamically new behaviors in the Andreev process $\cite{MRA,HHH}$ and superconducting Andreev states $\cite{ZM}$ at the interface of a normal-superconductor.

In addition above dynamical peculiar properties of $MoS_2$, its layered structure, chemical stability, and relatively high mobility (room temperature mobility over $200\; cm^2/Vs$) can make it potentially a useful material for electronics applications $\cite{RR,ZY,WK,BC,MH}$. The spin-valve effect in proximity-induced ferromagnetic $MoS_2$ may result in a valley-spin-resolved conductance $\cite{MA,MZA,KG}$, which is considered as an essential feature for valleytronic devices $\cite{DL,MHS,ZD}$. Analogous to graphene, there is a valley index $\tau=\pm 1$, which is robust against scattering by smooth deformations and long wavelength photons due to a large valley splitting. In this paper, we analytically investigate the signature of $p_x$-wave and chiral $p_x+ip_y$-wave symmetry in low-energy dispersions of Dirac-like charge carriers of ML-MDS. Electron-hole exchange at the interface of a normal-superconductor in quasiparticle excitations below superconducting gap leading to AR, and Andreev bound state (ABS) between two superconducting regions separated by a weak-link normal section can be affected by the off-diagonal elements of Bogoliubov-de Gennes gap matrix. The spin-triplet superconducting quasiparticle excitations in $MoS_2$ are found to play a crucial role in AR process, since the superconducting gap can be renormalized by electron-hole wave vector or as well chemical potential. Furthermore, the AR process is believed to be spin-valley polarized due to the valley-contrasting spin-splitting in the valence band and, indeed, depend on the magnitude of the chemical potential. Remarkably, the electron-hole difference mass $\alpha$ and topological $\beta$ terms $\cite{RM}$ are explicitly taken into the Dirac-Bogoliubov-de Gennes (DBdG) $MoS_2$ Hamiltonian, and thier contribution in superconductor electron-hole wavevector $k_{s}^{e(h)}$ gives rise to significantly change the resulting AR and ABS.

This paper is organized as follows. Section 2 is devoted to the analytical solutions of the ML-MDS DBdG equation with $p$-wave order parameter in order to obtain the exact expressions of dispersion energy, and corresponding Dirac spinors. We also represent the normal and Andreev reflection coefficients and Andreev bound state in the related NS and SNS junctions, respectively. The numerical results of subgap conductance and Josephson current are presented in Sec. 3 with a discussion of the main characteristics of systems. Finally, a brief conclusion is given in Sec. 4.

\section{THEORETICAL FORMALISM}
\subsection{Effective Hamiltonian}

The Brillouin zone of monolayer $MoS_2$ is hexagonal and around the edges of this zone, the low energy fermionic excitations behave as massive Dirac particles. The starting point for study their behavior is the tight-binding Hamiltonian. In addition to symmetry of the lattice, it is essential to consider the local atomic orbital symmetries. The minimum of conduction band is mainly formed from $d_{z^{2}}$ orbitals, and valence band maximum is constructed from orbitals $d_{x^{2}-y^{2}},d_{xy}$ of $Mo$ atom with mixing from $p_{x},p_{y}$ orbitals of $S$ atom in both cases. The full Hamiltonian for an arbitrary electron, labeled by the integer $\sigma$, that is included the possibility for symmetry adapted states and nearest neighbor hopping terms takes the form: 
\begin{equation}
H=\sum_{\sigma ij}\left[-\mu^{a}_{ij}a^{\dagger}_{\sigma i}a_{\sigma j}-\mu^{b}_{ij}b^{\dagger}_{\sigma i}b_{\sigma j}-\mu^{b^{'}}_{ij}b^{'\dagger}_{\sigma i}b^{'}_{\sigma j}\right]+\sum_{\left\langle \sigma\rho\right\rangle,ij}t_{\sigma\rho,ij}a^{\dagger}_{\sigma i}(b_{\rho j}+b^{'}_{\rho j})+H.c.
\end{equation}
Here $a$ and $b(b^{'})$ indicate the second quantized fermion operators on the $Mo$ and $S$ atoms in the up(down) layer, respectively. The indices $i$ and $j$ show the orbital degree of freedom labeled as $\left\{1,2,3\right\}\equiv\left\{d_{z^{2}},d_{x^{2}-y^{2}}+id_{xy},d_{x^{2}-y^{2}}-id_{xy}\right\}$ and $\{1^{'},2^{'}\}\equiv\left\{p_{x}+ip_{y},p_{x}-ip_{y}\right\}$ for $Mo$ and $S$ atoms, subsequently. Therefore the matrices $\mu^{a},\mu^{b},\mu^{b^{'}}$ and $t_{\sigma\rho,ij}$ are responsible for the on-site energies of $Mo$ and $S$ atoms and hopping between different neighboring sites in the space of different orbitals, respectively. Introducing the Slater-Koster method, the Hamiltonian density and overlap can be obtained $\cite{SK}$. To complete tight-binding Hamiltonian, it is necessary to add spin-orbit interaction in the model which causes spin-valley coupling in the valence band of ML-MDS. The most important contribution of this interaction is relevant to heavy metal $Mo$ atoms. We define the Fourier transform of fermionic field operators according to
$$
a_{\sigma is}=\sum_{\mathbf{k}}a_{\sigma ks}e^{i\mathbf{k}\cdot \mathbf{r_{\sigma}}},
$$
and similarly for $b_{\sigma i}$ and $b^{'}_{\sigma i}$. The spin-up and spin-down is labeled by $s=\pm 1$. The Hamiltonian can be written in momentum space $\hhh=\sum_{\mathbf{ks}}\psi^{\dagger}_{\mathbf{k}}H\psi_{\mathbf{k}}$, where $\psi_{\mathbf{k}}=\left(a_{ks1},a_{ks2},a_{ks3},b_{ks1^{'}},b_{ks2^{'}},b^{'}_{ks1^{'}},b^{'}_{ks2^{'}}\right)$.
This tight-binding model leads to calculate the effective electron and hole masses, energy gap, and valence band edges. This model was firstly presented by Di Xiao et al. $\cite{DL}$, and modified by Rostami et al. $\cite{RM}$ to obtain the explicit form of effective Hamiltonian for ML-MDS. Using the Lowdin partitioning method $\cite{W}$, the final result for effective low-energy two band continuum Hamiltonian governing the conduction and valence bands around the $K$ and $K^{'}$ points reads:
\begin{equation}
H=\hbar v_{F}\mathbf{k}\cdot\bsigma_{\tau}+\Delta\sigma_{z}+\lambda s\tau(\frac{1-\sigma_{z}}{2})+\frac{\hbar^{2}\left|k\right|^{2}}{2m_{0}}(\frac{\alpha}{2}+\frac{\beta}{2}\sigma_{z}),
\end{equation}
where $\bsigma_{\tau}=(\tau\sigma_{x},\sigma_{y})$ are the Pauli matrices. The valley index $\tau=\pm 1$ denotes the $K$ and $K'$ valleys. $\Delta$ is the direct band gap, $\lambda\approx 0.08\:eV$ $\cite{ZYY}$ and $v_{F}\approx 0.53\times 10^{6}\:ms^{-1}$ denote the spin-orbit coupling and Fermi velocity, respectively. The bare electron mass is $m_0$, and two numeric topological parameters are evaluated by $\alpha=m_{0}/m_{+}$ and $\beta=m_{0}/m_{-}-8m_{0}v^{2}_{F}/(\Delta-\lambda)$, where $m_{\pm}=m_{e}m_{h}/(m_{h}\pm m_{e})$. These band parameters originated from the difference between electron and hole masses ($\alpha$) and topological characteristics ($\beta$) have the values $0.43$ and $2.21$ $\cite{HPE}$, respectively. 

In proximity-induced superconducting $MoS_{2}$ case, the superconducting generalization of modified Dirac Hamiltonian can be written on the ML-MDS, including the pairing potential. In the relativistic case, Cooper pairing takes place between single particle states, which are constructed by application of the time reversal, parity and charge conjugation operators. The possible pairing states can be characterized as singlet or triplet order parameters. The superconducting order parameter is given by a function of spin $(s,s^{'})$ and momentum $(\mathbf{k},-\mathbf{k})$. Obviously, the spatial part of triplet order parameter is an odd function under exchange of the two particles, while the spin part is even. For a spin triplet symmetry the order parameter is expressed using the $d$-vector as:
\begin{equation}
\Delta_{s,s^{'}}(\mathbf{k})=(\mathbf{d}(\mathbf{k})\cdot\btau)i\tau_{y},
\end{equation}
where $\mathbf{d}(\mathbf{k})$ and $\btau$ and $\tau_y$ are an odd-parity function of $\mathbf{k}$ and Pauli matrices, which describe the real electron spin, respectively. The direction of the $d$-vector is perpendicular to the total spin $S$ of a Cooper pair. This order symmetry is off-diagonal, which differs from our previous work $\cite{HHH}$. Without loss of generality, let us consider the case of $\mathbf{d}(\mathbf{k})=\Delta_S(\mathbf{k})\hat{z}$, which means $\mathbf{k}\bot S$. The symmetry of the lattice plays, of course, a central role for $\mathbf{k}$-dependency of $\mathbf{d}$.  In ML-MDS, the promising pairing symmetry is $p_x$ and chiral  $p_{x}+ip_{y}$-wave $\cite{YLM}$. The superconducting potential interacting with lattice is written in the band picture as follows:
$$
H_{S}=\sum_{\left\langle \sigma\rho\right\rangle ij}\sum_{k}\sum_{s}V^{s}_{k}a^{\dagger}_{\sigma i}(b^{\dagger}_{\rho j}+b^{'\dagger}_{\rho j})+H.c.,
$$
where $V^{s}_{k}$ parameterizes the interaction strength of electrons, which can be determined according to the point group symmetry of $MoS_2$. Using the Taylor expansion around Dirac points, the low-energy limit of $H_S$ in the presence of ML-MDS can be calculated. Consequently, the corresponding DBdG Hamiltonian then reads:
\begin{equation}
\hhh=\left(\begin{array}{cccc}
A&v_{F}\hbar(\tau k_{x}-ik_{y})&0&\Delta_{S}(\mathbf{k})e^{i\varphi}\\
v_{F}\hbar(\tau k_{x}+ik_{y})&B&\Delta_{S}(\mathbf{k})e^{i\varphi}&0\\
0&\Delta_{S}^{\ast}(\mathbf{k})e^{-i\varphi}&-A&-v_{F}\hbar(\tau k_{x}-ik_{y})\\
\Delta_{S}^{\ast}(\mathbf{k})e^{-i\varphi}&0&-v_{F}\hbar(\tau k_{x}+ik_{y})&-B
\end{array}\right),
\end{equation}
where $A=\Delta+\frac{\hbar^{2}\left|k\right|^{2}}{2m_{0}}(\frac{\alpha}{2}+\frac{\beta}{2})-E_{F}+U(x)$ and $B=-\Delta+2\lambda s\tau+\frac{\hbar^{2}\left|k\right|^{2}}{2m_{0}}(\frac{\alpha}{2}-\frac{\beta}{2})-E_{F}+U(x)$. The globally broken $U(1)$ symmetry in the superconductor is characterized with phase $\varphi$. The electrostatic potential $U(x)$ gives the relative shift of Fermi energy ($E_{F}$) as $\mu_{n,s}=E_{F}-U(x)$, which denotes the chemical potential in N or S region. This matrix is Hermitian and may be diagonalized to yield the energy eigenvalues. Diagonalizing Eq. (4) produces the following energy-momentum quartic equation:
$$
\epsilon^{4}-2\epsilon^{2}\left(\frac{A^{2}+B^{2}}{2}+v^{2}_{F}\hbar^{2}\left|k_{s}\right|^{2}+\left|\Delta_S\right|^{2}\right)+\left(AB-v^{2}_{F}\hbar^{2}\left|k_{s}\right|^{2}\right)^{2}+
$$
$$
\left|\Delta_S\right|^{2}\left(2AB+2v^{2}_{F}\hbar^{2}\left|k_{s}\right|^{2}+\left|\Delta_S\right|^{2}\right)=0.
$$
The dispersion relation of DBdG for electron-hole excitations is found by:
\begin{equation}
\epsilon=\xi\sqrt{\left(\frac{A+B}{2}+\upsilon\sqrt{\frac{(A-B)^{2}}{4}+v^{2}_{F}\hbar^{2}\left|k_{s}\right|^{2}+\left|\Delta_S\right|^{2}(1-\eta^{2})}\right)^{2}+\left|\Delta_S\right|^{2}\eta^{2}},
\end{equation}
where $\eta^{2}=1-(\frac{A-B}{A+B})^{2}$. The parameter $\xi=\pm 1$ denotes the electron-like and hole-like excitations, while $\upsilon=\pm 1$ distinguishes between the conduction and valence bands. Note that, the superconducting order parameter $\Delta_{S}(\mathbf{k})$ is renormalized by chemical potential $\mu_s$, and also it appears as an ordinary gap. This electron-hole superconducting excitation is qualitatively different from that obtained for conventional singlet superconductivity $\cite{HHH}$ , so that it seems to remain \textit{semi-gapless}. The mean-field conditions are satisfied as long as $\Delta_{S}\ll \mu_s$. So, at this stage, it is appropriate to insert this restriction, which will be used in throughout this paper. In this condition, the exact form of superconductor wavevector of quasiparticles in $MoS_2$ can be acquired from this eigenstates
$$
k_{s}=\frac{1}{v_{F}\hbar}(k_{0}+ik_{1})\; ; \ \  k_{0}=\sqrt{AB},
$$
where $k_1$ may be responsible to exponentially decaying. In particular, we retain the contribution of $\alpha$ and $\beta$ terms representing one of the essential physics of $MoS_{2}$. 
The Hamiltonian Eq. (4) can be solved to obtain the wave function for superconductor region. We calculate the relevant electron-hole Dirac spinor, that is only valid for the spin triplet superconductivity. The wavefunctions, which includes a contribution from both electron-like and hole-like quasiparticles are found as:
\begin{equation}
\psi^{e}_{S}=\left(\begin{array}{cc}
\zeta \beta_{1}\\
\zeta \beta_{1}e^{i\tau\theta_s}\\
e^{-i\gamma_{e}}e^{-i\varphi}e^{i\tau\theta_s}\\
e^{-i\gamma_{e}}e^{-i\varphi}
\end{array}\right)e^{i(\tau k_{sx}x+k_{y}y)}, \ \ \psi^{h}_{S}=\left(\begin{array}{cc}
\zeta \beta_{2}\\
-\zeta \beta_{2}e^{-i\tau\theta_s}\\
-e^{-i\gamma_{h}}e^{-i\varphi}e^{-i\tau\theta_s}\\
e^{-i\gamma_{h}}e^{-i\varphi}
\end{array}\right)e^{i(-\tau k_{sx}x+k_{y}y)},
\end{equation}
where we define
$$
\beta_{1(2)}=-\frac{\epsilon}{\left|\Delta_S\right|}-(+)\sqrt{\frac{\epsilon^{2}}{\left|\Delta_S\right|^{2}}-\eta^{2}} ,\ \  \zeta=\frac{A+B}{2\sqrt{AB}} ,\ \ e^{i\gamma_{e}}=\frac{\Delta_S(\mathbf{k})}{\left|\Delta_S(\mathbf{k})\right|}.
$$
In analogous relation for $e^{i\gamma_{h}}$, it only needs to change the angle of incidence $\theta_{s}\rightarrow \pi-\theta_{s} $ in the order parameter.

\subsection{Scattering process in NS junction}

In this section, we will focus on the effect of induced spin-triplet $p$-wave symmetry on the transport properties of NS structure deposited on top of a ML-MDS. To this end, we assume that the superconducting region is located at $x>0$ and employ a scattering process follows from the Blonder-Tinkham-Klapwijk (BTK) formula $\cite{BTK}$. In the normal region ($x<0$), we have $\Delta_{S}(\mathbf{k})=0$, and an incident electron with energy excitation $\epsilon_{N}$ and transverse wave vector $k_{y}$ has two possible fates upon scattering, reflection as an electron or Andreev reflection as a hole. The Andreev reflection is the process that determines the conductance of the interface at bias voltage below the superconducting gap, because the incident electron can not transmit into the superconductor. We know that the superconducting pair potential couples an electron of spin $s$ from the valley $\tau$ with a hole of spin $-s$ from the valley $-\tau$. This spin-valley coupling is considered in all the next calculations.
At bias voltage above the superconducting gap, Andreev reflection is obviously suppressed, since direct tunneling into the superconductor is now possible. In fact, the unusual electronic properties of graphene leads to have the possibility of specular Andreev reflection. At the similar  ML-MDS-based interface, retro Andreev reflection can happen for subgap energies, since the direct band gap is too large to occur the specular Andreev reflection.

Having established the states that participate in the scattering, the total wave function for a right-moving electron with angle of incidence $\theta_{e}$, a left-moving electron by the substitution $\theta_{e}\rightarrow \pi-\theta_{e}$ and a left-moving hole by angle of reflection $\theta_{h}$ may then be written as
\begin{multline}
\psi_{N}=e^{ik_{y}y}\left(\frac{1}{\sqrt{\CN_{e}}}\left[1,\tau e^{i\tau\theta_{e}} A_{e},0,0\right]^{T}
e^{i\tau k^e_{x}x}+\right.\\
\left.\frac{r}{\sqrt{\CN_{e}}}\left[1,-\tau e^{-i\tau\theta_{e}} A_{e},0,0\right]^{T}e^{-i\tau k^e_{x}x}
+\frac{r_{A}}{\sqrt{\CN_{h}}}\left[0,0,1,\tau e^{-i\tau\theta_{h}} A_{h}\right]^{T}e^{i\tau k^h_{x}x}\right),
\end{multline}
where we define $A_{e(h)}=\hbar v_{F} \left|k^{e(h)}\right|/((-)\epsilon_{N}-B)$. Here, $r$ and $r_{A}$ are the normal and Andreev scattering coefficients, respectively. The normalization factor $\CN_{e(h)}$ ensure that the quasiparticle current density of states is the same. The charge and current density of quasiparticles may be defined by nonrelativistic and relativistic terms based on the Lorentz covariance continuity equation. Using the modified Dirac Hamiltonian, the normalization factors result in:
$$
\CN_{e(h)}=A_{e(h)}\cos(\tau\theta_{e(h)})+\frac{\tau\hbar \left|k^{e(h)}\right|}{4m_{0}v_{F}}\left((\alpha+\beta)+A^{2}_{e(h)}(\alpha-\beta)\right)\cos(\tau\theta_{e(h)}).
$$
Diagonalizing the Hamiltonian in the normal region provides the energy eigenvalue for electron (relative to the Fermi energy) $\epsilon_{N}=\left|\frac{A+B}{2}\pm\sqrt{\left(\frac{A-B}{2}\right)^2+v^{2}_{F}\hbar^{2}\left|k\right|^{2}}\right|$. The $\pm$ refers to the excitations in the conduction and valence bands. The nonconserved component of the momentum $k_{x}=|k|\cos{\theta_{e}}$ can be acquired from this eigenstates. It is instructive to consider the effect of Fermi vector mismatch on Andreev reflection amplitude. To explore how the Fermi vector mismatch influences the scattering processes, the Fermi momentum in the normal and superconducting part of system may be controlled by means of a chemical potential. The strategy for calculating the scattering coefficients in the junction is to match the wave functions in normal and superconductor regions, $\psi_{N}=\psi_{S}$ at $x=0$, where $\psi_{S}=t^{e}\psi^{e}_{S}+t^{h}\psi^{h}_{S}$. The coefficients $t^{e}$ and $t^{h}$ correspond to the transmission of electron and hole quasiparticle, respectively. The four scattering amplitudes together form the system of equations where we define
$$
\Xi S=\Psi,\ \ \ S=\left[r,r_{A},t^{e},t^{h}\right]^{T},\ \ \ \Psi=\left[1,\tau e^{i\tau\theta_{e}} A_{e},0,0\right]^{T},
$$
and the analytical expression of 4$\times$4 matrix $\Xi$ (given explicitly in the Appendix A) is obtained by solving the boundary condition. We find the following solution for the normal and Andreev reflection coefficients
\begin{equation}
r_{A}=\sqrt{\frac{\CN_{h}}{\CN_{e}}}\frac{2\tau A_{e} \cos{\theta_{e}}}{\eta_{1}\eta_{4}-\eta_{2}\eta_{3}}\left[\eta_{4}e^{i\tau\theta_s}e^{-i\gamma_{e}}-\eta_{3}e^{-i\tau\theta_s}e^{-i\gamma_{h}}\right],
\end{equation}
\begin{equation}
r=\frac{2 \tau A_{e} \cos{\theta_{e}}}{\eta_{1}\eta_{4}-\eta_{2}\eta_{3}}\left[\eta_{4}\beta_{1}+\eta_{3}\beta_{2}\right]-1,
\end{equation}
where we have introduced
$$
\eta_{1(2)}=\zeta \beta_{1(2)}\left[(-)\tau e^{-i\tau\theta_{e}} A_{e}+e^{(-)i\tau\theta_s}\right],
$$
$$
\eta_{3(4)}=e^{-i\gamma_{e(h)}}\left[\tau e^{-i\tau\theta_{h}} A_{h}e^{(-)i\tau\theta_s}-(+)1\right].
$$
We will verify that the dispersion energy for triplet pairing symmetry have anomalous properties. Thus, the results for scattering amplitudes exhibit qualitatively distinct behavior. One obviously verifies that Andreev reflection may be slightly suppressed in the presence of triplet superconducting gap. Here, the order parameter may written as $\Delta_{S}(\mathbf{k})=\Delta_{S}\cos{\theta_{s}}$ for $p_x$-wave and $\Delta_{S}e^{i\theta_{s}}$ for chiral $p_x+ip_y$-wave symmetries.

In follows, according to the BTK formalism, we can calculate the tunneling conductance
\begin{equation}
G(eV)=\sum_{s,\tau=\pm 1} G^{s,\tau}_{0}\int^{\theta_{c}}_{0}\left(1-\left|r\right|^{2}+\left|r_{A}\right|^{2}\right)\cos{\theta_{e}}d\theta_{e},
\end{equation}
where $G^{s,\tau}_{0}=e^{2}N_{s,\tau}(eV)/h$ is the ballistic conductance of spin and valley-dependent transverse modes $N_{s,\tau}=kw/\pi$ in a $MoS_{2}$ sheet of width $w$, and $eV$ denotes the bias voltage. The upper limit of integration in Eq. (10) needs to determine based on the fact that the incidence angle of electron-hole at the interface may be less than $\pi/2$ in the two N and S regions.

\subsection{SNS Josephson junction}

We now proceed to study the Andreev bound states and current-phase relation in a ML-MDS Josephson junction, where the normal region is extended from $x=0$ to $x=d$. ML-MDS is covered by superconducting electrodes in the regions $x>d$ and $x<0$. At a superconductor surface, electrons(holes) are reflected as holes(electrons), and their energy lies within the superconducting energy gap. In Josephson junction, the overlap of the wavefunctions of the surface states builds up the ABS, which play an essential role in Josephson transport. The electron and hole wave function at a SN interface are globally coupled by a scattering matrix.

Firstly, we obtain the energy spectrum for the ABSs in the normal region by matching the wave functions at the two interfaces, i.e. $\psi^{L}_{S}=\psi_{N}$ at $x=0$, and $\psi_{N}=\psi^{R}_{S}$ at $x=d$. The left and right superconductor wavefunctions $\psi^L_S$ and $\psi^R_S$ obey generally from the Eq. (6) for left-right moving electron and hole. The wave function in N region is a superposition of right and left moving electrons and holes due to Andreev reflection process. Proceeding the same procedure in the previous section, we can find the following system of equations
\begin{equation}
\Xi' S=0 \ ; \ \ \ S=\left[t^{e}_{L},t^{h}_{L},a,b,c,d,t^{e}_{R},t^{h}_{R}\right]^{T},
\end{equation}
where $\Xi'$ is a 8$\times$8 matrix of coefficient given explicitly in the Appendix B. Subsequently, the ABS carrying the Josephson current in the normal region can be obtain from nontrivial solution for the Eq. (11). These bound state energies are actually expressed in terms of macroscopic phase difference ($\Delta\varphi=\varphi_{R}-\varphi_{L}$) between the left and right superconducting sections:
\begin{equation}
\epsilon(\Delta\varphi)=\eta\Delta_S(\theta_s)\sqrt{\frac{1}{2}\left[1-\frac{\Gamma(\Delta\varphi)}{\Omega}\right]},
\end{equation}
where we have defined the auxiliary quantities
$$
\Gamma(\Delta\varphi)=-e^{-i\gamma_{e}}e^{-i\gamma_{h}}\left[2\cos^2{\theta_s}\cos^2{\theta_{e}}\cos{\Delta\varphi}+\sin^2{k_{x}d}\left(1-\cos{2\theta_s}-4\sin{\theta_e}\sin{\theta_s}+2\sin^2{\theta_{e}}\right)\right],
$$
$$
\Omega=2\left[\sin^2{k_{x}d}\left(A_{e}\sin{\theta_e}\sin{\theta_s}-1\right)^2+\cos^2{k_{x}d}\left(A_{e}\cos{\theta_e}\cos{\theta_s}\right)^2\right].
$$
The analytical progress can be made in the limit of thin and strong barrier, which the barrier strength parameter is then defined as $Z=k_{x}d$. Otherwise, the results of analytical calculations can be presented as a function of length of junction. In short junction limit, the length of the junction is smaller than the superconducting coherence length $\xi=\hbar v_{F}/\Delta_S$.

The Josephson current-phase relation may now be computed via the standard expression
\begin{equation}
\ii/\ii_{0}=\int^{\pi/2}_{-\pi/2}\:d\theta_s\cos{\theta_s}\tanh{\left(\frac{\epsilon(\Delta\varphi)}{2K_{B}T}\right)}\frac{d\epsilon(\Delta\varphi)}{d\Delta\varphi}.
\end{equation}
Here, $K_{B}$ and $T$ are the Boltzmann constant and temperature, respectively. $\ii_{0}=4e/\hbar$ denotes the normalized current. Note that the factor of $4$ in the normalized current is appeared due to the spin and valley degeneracy.

\section{RESULTS AND DISCUSSION}
\subsection{Andreev process}

In this section, we will analyze in detail the dynamical transport properties of hybrid structure deposited on top of a ML-MDS in order to investigate how the triplet $p$-wave superconductivity can affect the Andreev process and resulting subgap conductance. The $MoS_2$ superconducting electron-hole excitations of Eq. (5) is qualitatively different from that in $s$-wave case (see, Eq. (4) in Ref. $\cite{HHH}$). Particularly, we demonstrate this quantity in Figs. 1(a) and (b), where the valence band energy with fixed spin polarization represents the superconducting subgap. The Fermi wave vector and also electron-hole branches (the electron branch for $|k|>k_F$, and the hole branch for $|k|<k_F$) are shifted by the essential features of monolayer $MoS_2$ Hamiltonian. The magnitude of superconducting order parameter $\Delta_S$ is now renormalized by the coefficient $\eta^2$, which contains the dynamical characteristic of ML-MDS, and also chemical potential. To clearly see this feature, we take the net value of pair potential $\Delta_S=0.1\;eV$, although we will not further need to use this value in the next calculations, since superconducting excitation energy $\epsilon$ may necessarily be normalized by pair potential $\Delta_S$.

For actual case of ML-MDS, which the asymmetry mass term $\alpha$ originated from electron-hole difference mass and topological $\beta$ term are taken into Hamiltonian, the superconductor subgap is reduced to a value $\sim 0.06\;eV$ (see the black line in Fig. 1(a)). This, therefore, can lead to suppress the AR. It is seen that the strong SOC in $MoS_2$ may enhance the AR, while the topological term $\beta$ will give rise to more suppression of AR at the NS interface. The mass-related parameter $\alpha$ has no effect on the valence band excitation. These features are presented in Fig. 1(b).

Based on above triplet superconductor valence band energy behavior, the resulted Andreev and normal reflections are demonstrated in Fig. 2(a) and (b) at zero energy $\epsilon/\Delta_S=0$, where N region is necessarily $p$-doped. The S region is taken to be $p$-doped, $\mu_n=\mu_s=-0.96\; eV$ in Fig. 2(a). In this condition, the suppression of AR is significant, so that the maximum suppressed AR happens in the absence of SOC, and as resulted from excitation energy curve, the AR becomes a considerable value for the case when $\beta=0$ due to increasing the subgap energy. Furthermore, the probability density of reflections is conserved, i. e. $|r_A|^2+|r|^2=1$, since we are in zero bias energy.
To evaluate the $n$-doped S region effect, we plot the $\mu_s$ dependence of the normal and AR probabilities versus electron angle of incidence, as shown in Fig. 2(b), where the created Cooper pair in S region will be in the conduction band of superconductor $MoS_2$. In this condition, we have Fermi wavevector mismatch (FVM) between reflected hole in N region and corresponding transmitted electron to the S region, which means that $\mu_s\gg\Delta_S$. We see that by increasing the chemical potential in S region the magnitude of AR probability has a noticeable value for most incidence angles upto $\pi/2$, since we are in $n$-doped condition. Moreover, the normal and Andreev reflections exhibit fundamentally new behavior when the bias energy is a non-zero value $\epsilon/\Delta_S\neq 0$. 
To more clarify this situation, we calculate the Andreev resonance states at the interface of NS junction by the fact that the normal reflection coefficient may be equal to zero, which yields the following solution: 
$$
\epsilon=\eta\Delta_S(\theta_s)sgn\left[\tan\left(\frac{1}{2i}\ln\left(-\omega_{2}/\omega_{1}\right)\right)\right]\sqrt{\frac{\omega^{2}_{4}}{\omega^{2}_{4}-\omega^{2}_{3}}} ,
$$
where we define
$$
\omega_{1(2)}=\eta_{4(3)}\left(2\tau A_{e}\cos{\theta_{e}}-\eta_{4(3)}e^{(-)i\tau\theta_s}e^{i\gamma_{h(e)}}\right),\ \ \ \omega_{3(4)}=\omega_{1}+(-)\omega_{2}.
$$
To explore how doping in superconductor region influences this energies, we plot it versus the electron incidence angle in N region. As seen in Fig. 3, the slope of resonance energy curves around normal incidence angle varies very slightly with increasing the chemical potential of S region. Consequently, the resonance energy contribution to the supercurrent conductance becomes considerable around $\theta_{e}=0$, and the height of the zero-bias conductance is enhanced by increasing $\mu_{s}$. In the $p$-doped case, the dispersion is weak, so the bound state energy is close to zero $\epsilon=0$. Therefore, the doping of S region has the important effect on the resonance states. It is seen that the behavior of resonance states around large angles of incidence ($\theta_e>0.1\pi$) in the $p$-doped case is qualitatively different from the $n$-doped case, so that, there is a zero slope of Andreev states at $\theta_e=\pi/2$, and the magnitude of energies becomes a maximum value.

Fig. 4 demonstrates the reflection probability curves in terms of bias energies below its normalized magnitude for $p$-doped S region. Note that, the normalized bias is reduced from a magnitude $\epsilon=1\Delta_S$ to $\sim 0.31\Delta_S$, when we take into the contribution of all dynamical parameters ($\lambda, \alpha, \beta$) of ML-MDS. However, the normal and Andreev reflections tend to a magnitude of unit and zero, respectively, when the angle of incidence goes to the $\pi/2$. We have no reflections probability conservation ($|r_A|^2+|r|^2\neq 1$) due to the non-zero bias energy. A sharp minimum peak is observed in AR at all incidence angles $0\leq\theta_e\leq\pi/2$, so that the position of dip can be displaced by the \textit{bias limitation} coefficient $\eta$.

Importantly, in either $p$- or $n$-doped S region cases, the perfect AR process occurs with unit value for normal incidence angle. The subgap conductance of NS structure resulted from Andreev process is plotted in Fig. 5(a). For chemical potential of N region $\mu_n=-0.96\; eV$, the suppression of AR results in a low conductance for $p$-doped S region ($\mu_s=-0.96\; eV$), whereas we find larger conductance for $n$-doped condition at the bias energies $0.15<\epsilon/\Delta_s<0.25$. Increasing the doping $\mu_s\geq 3\; eV$ affects the superconducting subgap, and conductance curve strongly descends in high biases. A sharp conductance dip close to zero relative to coefficient $\eta$ is obtained for all magnitude of chemical potential in S region. The influence of dynamical characteristics of ML-MDS in conductance is presented in Fig. 5(b). As expected, the absence of topological term $\beta$ results in higher conductance, and mass-related term $\alpha$ has no substantial effect on the transport properties of structure. Also, in Fig. 5(b), one can see the effect of SOC, where the zero dip of conductance appeared in $p$- and $n$-doped S region cases ($\mu_n=-1\; eV, \mu_s=-1,\; 3\; eV$) is displaced when the SOC vanishes $\lambda=0\; eV$. Indeed, spin-valley coupling in the non-degenerate $K$ and $K'$ valleys and also valence band spin-splitting resulted from strong SOC gives rise to decrease the conductance in $p$-doped case. When we neglect the above effects, i.e. the SOC to be zero the conductance increases in the case of $n$-doped S region.

\subsection{Andreev bound state}

In order to study the effect of triplet superconductivity on the supercurrent passing through ML-MDS SNS Josephson junction, we proceed to investigate the ABS given by the expression of Eq. (12). In this equation, the electron angle of incidence in the normal region separated the two superconducting regions yields $\theta_{e}=\sin^{-1}(k_s/k\sin\theta_s)$, which corresponds to lower chemical potential (we may have, in this case, the condition $\mu_n\approx\mu_s$, due to the existence of direct energy gap in low-energy excitations of ML-MDS). In the case of a thin and strong barrier for N region ($\mu_n\gg\mu_s$), one can take the $\theta_{e}$ to be zero and, therefore, a barrier strength parameter $Z$ can be realized. Exploring the obtained bound state of Eq. (12) reveals the fact that the slope of ABS curves (we do not present the related figure) at $\Delta\varphi=0$ and $2\pi$. This is not a correct physical result indicating the non-zero (no finite) supercurrent in zero phase difference. Such situation can be understood from the fact that the zero energy states (ZESs) may be formed near the interfaces due to unconventional superconductivity, as shown in the similar situation in Ref. $\cite{KSY,YS}$. 
Therefore, in what follows, we have to derive the new energy states corresponding to ZESs, which, in general, can be given as:
\begin{equation}
\epsilon(\Delta\varphi)=\eta\Delta_S(\theta_s)\sqrt{\ee} \cos{\left(\frac{\Delta\varphi}{2}\right)},
\end{equation}
where
$$
\ee=\frac{e^{-i\gamma_{e}}e^{-i\gamma_{h}}\cos^2{\theta_s}\cos^2{\theta_{e}}}{\sin^2{(k_{x}d)}\left(A_{e}\sin{\theta_e}\sin{\theta_s}-1\right)^2+\cos^2{(k_{x}d)}\left(A_{e}\cos{\theta_e}\cos{\theta_s}\right)^2} .
$$
In fact, the $p_x$-wave symmetry satisfies the condition $\Delta_S(\theta_s)=-\Delta_S(\pi-\theta_s)$, which causes the existence of ZESs. Also, we have the similar condition for chiral $p_x+ip_y$-wave symmetry.

Figure 6 presents the ABS featuring ZESs for $p_x$-wave or chiral $p_x+ip_y$-wave symmetries, when $\mu_s=-1.1\; eV$ and $Z=0.5\pi$. As seen, the maximum slope of bound energies occurs in $\Delta\varphi=\pm\pi$, whereas $\Delta\varphi=0,\; \pm 2\pi$ results in flat energy curve, which may correspond to $0-\pi$ current-phase relation. By increasing the quasiparticle incidence angle, the energy states remain gapless. We find that flattening ABS in $p_x$-wave symmetry quickens rather than chiral $p_x+ip_y$-wave. Remarkably, the dynamical feature of $MoS_2$ affects significantly the ABS in a way that the effect of SOC ($\lambda\neq 0$) enhances and topological term $\beta\neq 0$ reduces the ABS. The mass asymmetry term $\alpha$ has no further effect. These features are demonstrated in Fig. 6. We also proceed to discuss the barrier strength and quasiparticle angle of incidence dependence of energy states formed in N region, and especially investigate how they vary from $p_x$-wave case to the chiral $p_x+ip_y$-wave symmetry. Figs. 7(a) and (b) show that the ABS corresponding to ZESs displays regular oscillations as a function of the barrier strength in the case of $\mu_s\gg\Delta_S$ and $\Delta\varphi=0$ or $2\pi$. The peaks of oscillations for both cases of order parameters are in $Z=n\pi$ ($n$ is integer), so that, by increasing the incidence angle the magnitude of peaks decreases in $p_x$-wave symmetry (Fig. 7(a)), while it is a constant value $\sim 0.47$ in chiral $p_x+ip_y$-wave (Fig. 7(b)).

Finally, we consider the resulting current-phase relationship for triplet order parameter featuring ZESs, in which the zero-temperature limit is assumed in all the following plots. In Figs. 8(a) and (b), we plot the phase difference and barrier strength dependences of Josephson current for $p_x$ and $p_x+ip_y$-wave symmetries when the $p$-doped S region is taken, $\mu_s=-1\; eV$. The abrupt crossover current at $\Delta\varphi=\pi$ can be understood by ZESs feature of ABS. From $p_{x}$-wave to chiral $p_x+ip_y$-wave, the magnitude of current decreases, and its sign is changed, which can be justified by a factor of $\cos{\theta_{s}}$ in the $p_x$-wave case. The current shows an oscillatory behavior as a function of barrier strength $Z$, and its maximum occurs in $Z=n\pi$, according to the energy bound states. In order to see the effect of doping on the critical current, we plot the width $d$ of N region dependence of critical current for $n$-doped S region and $\mu_n=-1.3\; eV$, in Fig. 9. Existence of FVM, in this case, causes to decline the maximum supercurrent with a $(d/\xi)^{-1}$ relationship. The case of $p$-doped S region $(\mu_s=-1.3\; eV)$, when the Fermi level lies between the spin-split valence subband of $MoS_2$ the Josephson current curve is presented to compare with the $n$-doped case (see the black line in Fig. 9).
Subsequently, the critical current exhibits oscillatory function in terms of width $d$, where the oscillation-amplitude is considerably small. Indeed, the absence of FVM between S and N regions is responsible to this behavior, which is shown in Fig. 10.

\section{CONCLUSION}

In summary, we have investigated the effect of proximity-induced spin triplet superconducting monolayer molybdenum disulfide on the transport properties of normal-superconductor and superconductor-normal-superconductor hybrid junctions. One of key findings of the present work is that the $MoS_2$ superconducting electron-hole excitations remain semi-gapless. This feature has led to suppress more or less the Andreev reflection in the case of $p$-doped S region, since the superconducting subgap weakens with a bias limitation coefficient $\eta$, which contains the dynamical characteristics of $MoS_2$ such as mass asymmetry, topological and spin-orbit coupling terms, and, of course, the chemical potential of each region. The perfect AR has been found at normal incidence to the interface with both $p$ and $n$-doped S region, and AR enhances in $n$-doped case. The resulting Andreev conductance for $p$-doped S region has a low value, while it considerably increases in the bias energies $0.15<\epsilon/\Delta_S<0.25$ for $n$-doped case. On the other hand, the Andreev bound state behavior as a function of phase difference between two superconducting sections in Josephson junction reveals the formation of zero energy states at the interfaces, which can be understood by the intrinsic feature of unconventional order parameters. We have obtained an abrupt crossover current-phase relation curve. The critical current has been found to strongly depend on the doping of S region. We have reported on critical current oscillations in terms of junction length in the absence of Fermi wavevector mismatch. In the case of thin and strong barrier, the supercurrent exhibits an oscillatory behavior versus barrier strength with a period of $Z=n\pi$.

  \renewcommand{\theequation}{A-\arabic{equation}}
  \setcounter{equation}{0}  
  \section*{APPENDIX A: Reflection scattering matrix}  
In this appendix, to complete the construction of scattering process in NS junction in detail, we introduce the 4$\times$4 matrix $\Xi$, which is obtained from the boundary condition at the interface for wavefunctions of N and S regions:  
\be
\Xi=\left(\begin{array}{cccc}
\CN^{-1/2}_{e}&0&-\zeta \beta_{1}&-\zeta \beta_{2}\\
-\CN^{-1/2}_{e}\tau e^{-i\tau\theta_{e}} A_{e}&0&-\zeta \beta_{1}e^{i\tau\theta_s}&\zeta \beta_{2}e^{-i\tau\theta_s}\\
0&\CN^{-1/2}_{h}&-e^{i\tau\theta_s}e^{-i\gamma_{e}}e^{-i\varphi}&e^{-i\tau\theta_s}e^{-i\gamma_{h}}e^{-i\varphi}\\
0&\CN^{-1/2}_{h}\tau e^{-i\tau\theta_{h}} A_{h}&-e^{-i\gamma_{e}}e^{-i\varphi}&-e^{-i\gamma_{h}}e^{-i\varphi}
\end{array}\right).
\ene
  
  \renewcommand{\theequation}{B-\arabic{equation}}
  \setcounter{equation}{0}  
  \section*{APPENDIX B: Josephson scattering matrix}  
  
Here, as described in the text, we give the 8$\times$8 matrix $\Xi'$ in the form of four 4$\times$4 matrices, which are used to calculate the Andreev energy bound states and corresponding Josephson supercurrent in the SNS junction:   

\begin{equation}
\Xi'=\left(\begin{array}{cc}
\ca_1&\ca_{2}\\
\ca_{3}&\ca_{4}
\end{array}\right),
\end{equation}
where we have
$$
\ca_{1}=\left(\begin{array}{cccc}
\zeta \beta_{1}&\zeta \beta_{2}&-1&-1\\
-\zeta \beta_{1}e^{-i\tau\theta_s}&\zeta \beta_{2}e^{i\tau\theta_s}&-\tau e^{i\tau\theta_{e}} A_{e}&\tau e^{-i\tau\theta_{e}} A_{e}\\
-e^{-i\gamma_{e}}e^{-i\varphi_{1}}e^{-i\tau\theta_s}&e^{-i\gamma_{h}}e^{-i\varphi_{1}}e^{i\tau\theta_s}&0&0\\
e^{-i\gamma_{e}}e^{-i\varphi_{1}}&e^{-i\gamma_{h}}e^{-i\varphi_{1}}&0&0
\end{array}\right);
$$
$$
\ca_{2}=\left(\begin{array}{cccc}
0&0&0&0\\
0&0&0&0\\
-1&-1&0&0\\
\tau e^{i\tau\theta_{h}} A_{h}&-\tau e^{-i\tau\theta_{h}} A_{h}&0&0
\end{array}\right);
$$
$$
\ca_{3}=\left(\begin{array}{cccc}
0&0&e^{i\tau k^{e}_{x}d}&e^{-i\tau k^{e}_{x}d}\\
0&0&\tau e^{i\tau\theta_{e}} A_{e}e^{i\tau k^{e}_{x}d}&-\tau e^{-i\tau\theta_{e}} A_{e}e^{-i\tau k^{e}_{x}d}\\
0&0&0&0\\
0&0&0&0
\end{array}\right);
$$
$$
\ca_{4}=\left(\begin{array}{cccc}
0&0&-\zeta \beta_{1}e^{i\tau k_{Sx}d}&-\zeta \beta_{2}e^{-i\tau k_{Sx}d}\\
0&0&-\zeta \beta_{1}e^{i\tau\theta_s}e^{i\tau k_{Sx}d}&\zeta \beta_{2}e^{-i\tau\theta_s}e^{-i\tau k_{Sx}d}\\
e^{-i\tau k^{h}_{x}d}&e^{i\tau k^{h}_{x}d}&-e^{-i\gamma_{e}}e^{-i\varphi_{2}}e^{i\tau\theta_s}e^{i\tau k_{Sx}d}&e^{-i\gamma_{h}}e^{-i\varphi_{2}}e^{-i\tau\theta_s}e^{-i\tau k_{Sx}d}\\
-\tau e^{i\tau\theta_{h}} A_{h}e^{-i\tau k^{h}_{x}d}&\tau e^{-i\tau\theta_{h}} A_{h}e^{i\tau k^{h}_{x}d}&-e^{-i\gamma_{e}}e^{-i\varphi_{2}}e^{i\tau k_{Sx}d}&-e^{-i\gamma_{h}}e^{-i\varphi_{2}}e^{-i\tau k_{Sx}d}
\end{array}\right).
$$

\newpage

\textbf{Figure captions}\\
\textbf{Figure 1(a), (b)} (color online) (a) The excitation spectra in superconductor ML-MDS, calculated from Eq. 5 when $\Delta_S=0.1\;eV$ and $\mu_s=-1.1\;eV$ . Green and blue lines indicate energy dispersion for $\lambda=0.08\;eV$, while red and black lines for $\lambda=0\;eV$. Green and red lines indicate excitations for $\alpha=\beta=0$. (b)  Effect of $\alpha$ and $\beta$ terms on the excitation spectra is indicated, separately.\\
\textbf{Figure 2(a), (b)}(color online) Probability of the normal and Andreev reflection as a function of the incidence angle for (a) $p$-doped S region ($\mu_n=\mu_s=-0.96\;eV$) and (b) $n$-doped S region ($\mu_n=0.96\;eV$). The plots in (a) show the results for different values of $\alpha,\beta,\lambda$ and (b) show the results for various values of superconducting chemical potential.\\
\textbf{Figure 3}(color online) The resonance energy as a function of the electron incident angle for several values of superconductor chemical potential in the $p$- and $n$-doped cases.\\
\textbf{Figure 4}(color online) Plot of the probability of the normal and Andreev reflection as function of bias voltage with $p$-doped S region while $\mu_n=\mu_s=-0.96\;eV$. The curves show the results for various angles of incidence.\\
\textbf{Figure 5(a), (b)}(color online) The tunneling conductance for several values of the superconducting chemical potential for a NS junction with $\mu_n=-0.96\;eV $. The plots in (a) show the results for different values of $\mu_s$ and effect of superconductor doping is indicated and (b) show the results for different values of $\alpha,\beta,\lambda$.\\ 
\textbf{Figure 6}(color online) Plot of the Andreev bound state energy versus phase difference in Josephson SNS junction. The role of the topological terms and spin-orbit coupling has been demonstrated. We have set $\mu_s=-1.1\;eV$ and $Z=0.5\;\pi$.\\
\textbf{Figure 7(a), (b)}(color online) Plot of the bound state energy versus barrier strength and superconductor angles of incidence for (a) $p_x$-wave and (b) $p_x+ip_y$-wave symmetry.\\
\textbf{Figure 8(a), (b)}(color online) Josephson current for (a) $p_x$-wave and (b) $p_x+ip_y$-wave symmetry as function of phase difference and barrier strength when $\mu_s=-1\;eV$.\\
\textbf{Figure 9}(color online) Plot of current-phase relation difference for $p_x$-wave symmetry with a FVM between the S and N regions. Also, we plot the length dependence of the critical current of Josephson junction for various values of superconductor chemical potential. We have set $\mu_n=-1.3\;eV$.\\
\textbf{Figure 10}(color online) Plot of the length dependence of the critical current for $p_x$-wave symmetry with no FVM between the S and N regions. We have set $\mu_n=\mu_s =-1.2\;eV $.\\

\newpage

\begin{figure}[p]
\epsfxsize=0.4 \textwidth
\begin{center}
\epsfbox{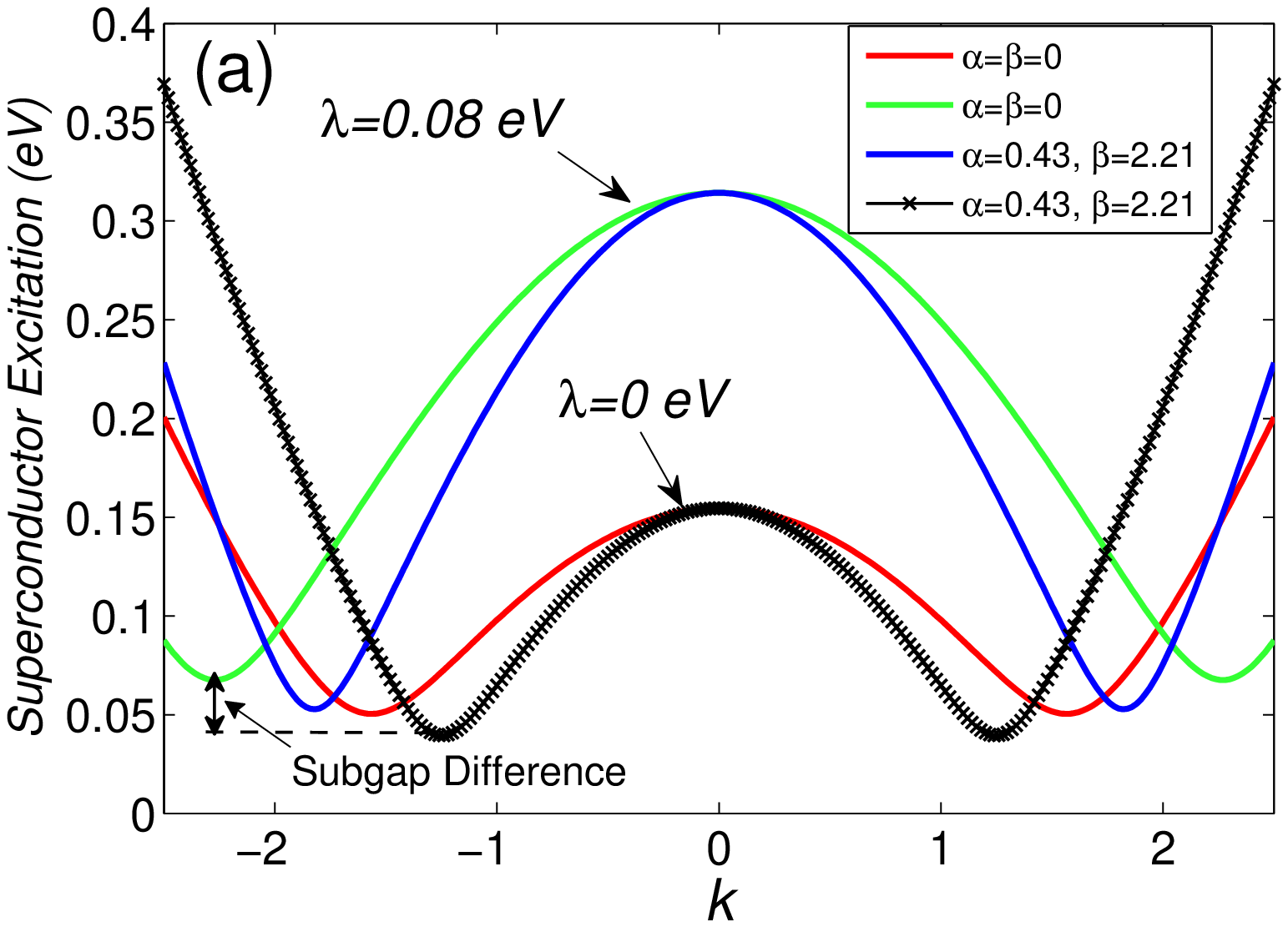}
\end{center}
\end{figure}

\begin{figure}[p]
\epsfxsize=0.4 \textwidth
\begin{center}
\epsfbox{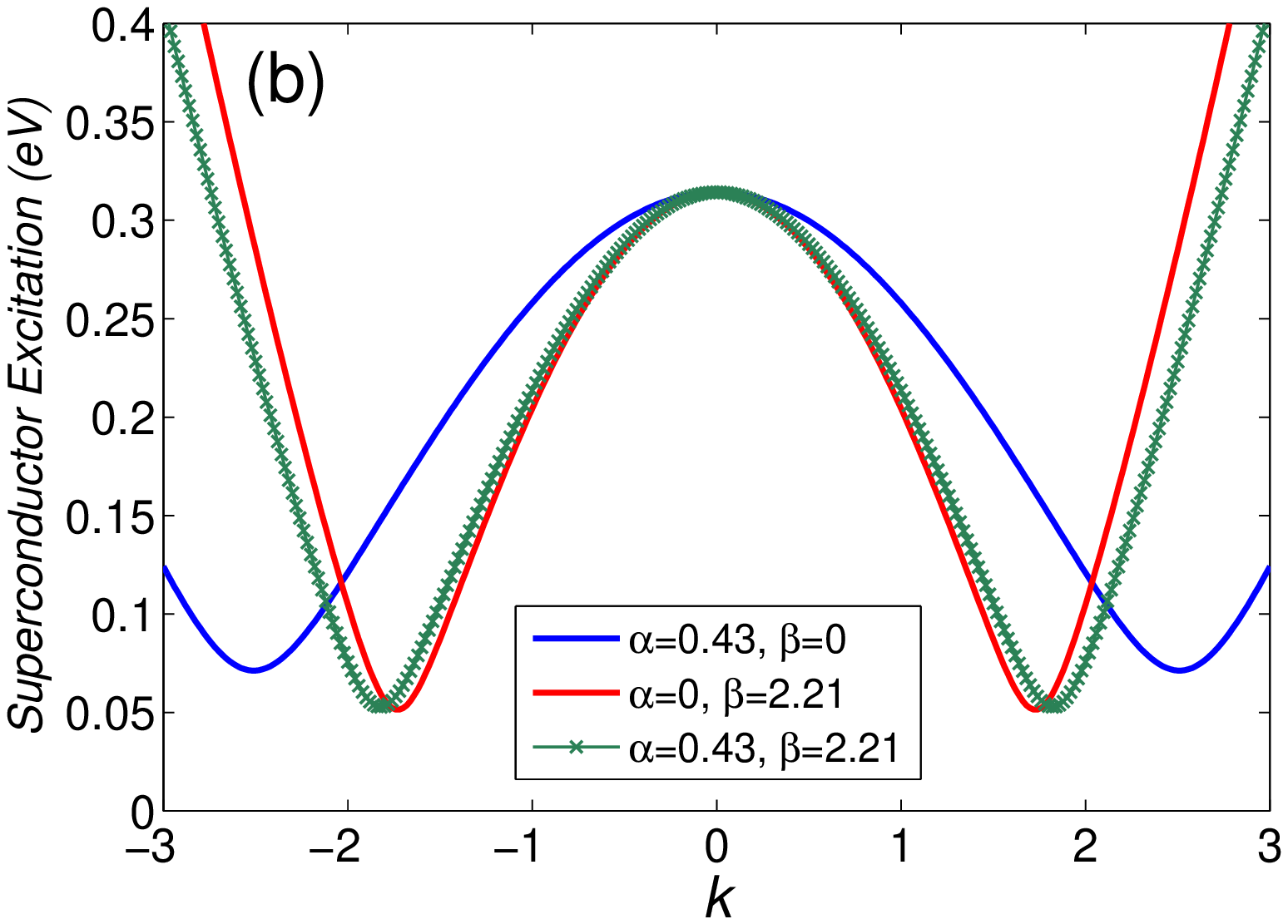}
\setcounter{figure}{0}
\caption{\footnotesize (a), (b)}
\end{center}
\end{figure}

\begin{figure}[p]
\epsfxsize=0.4 \textwidth
\begin{center}
\epsfbox{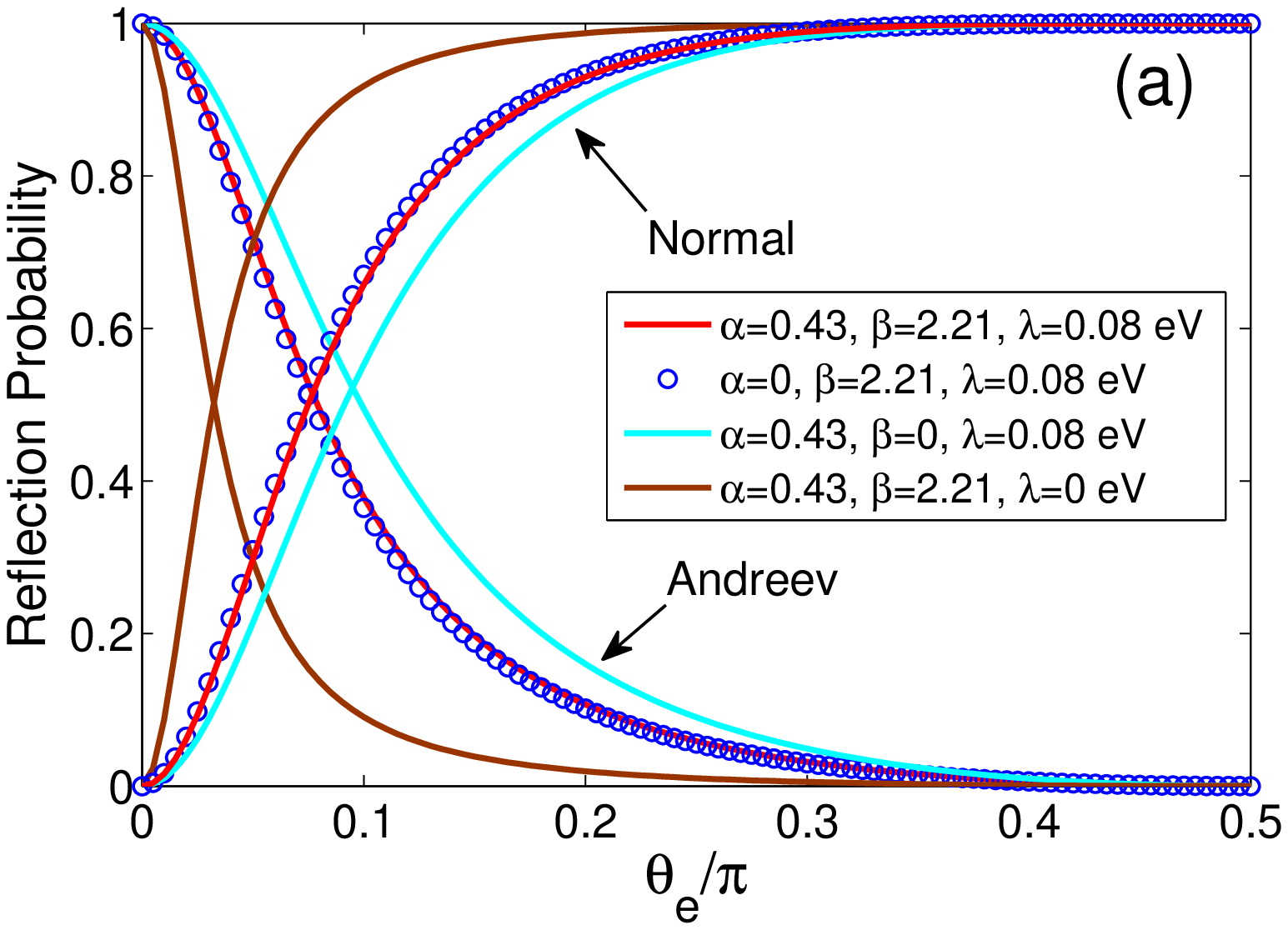}
\end{center}
\end{figure}

\begin{figure}[p]
\epsfxsize=0.4 \textwidth
\begin{center}
\epsfbox{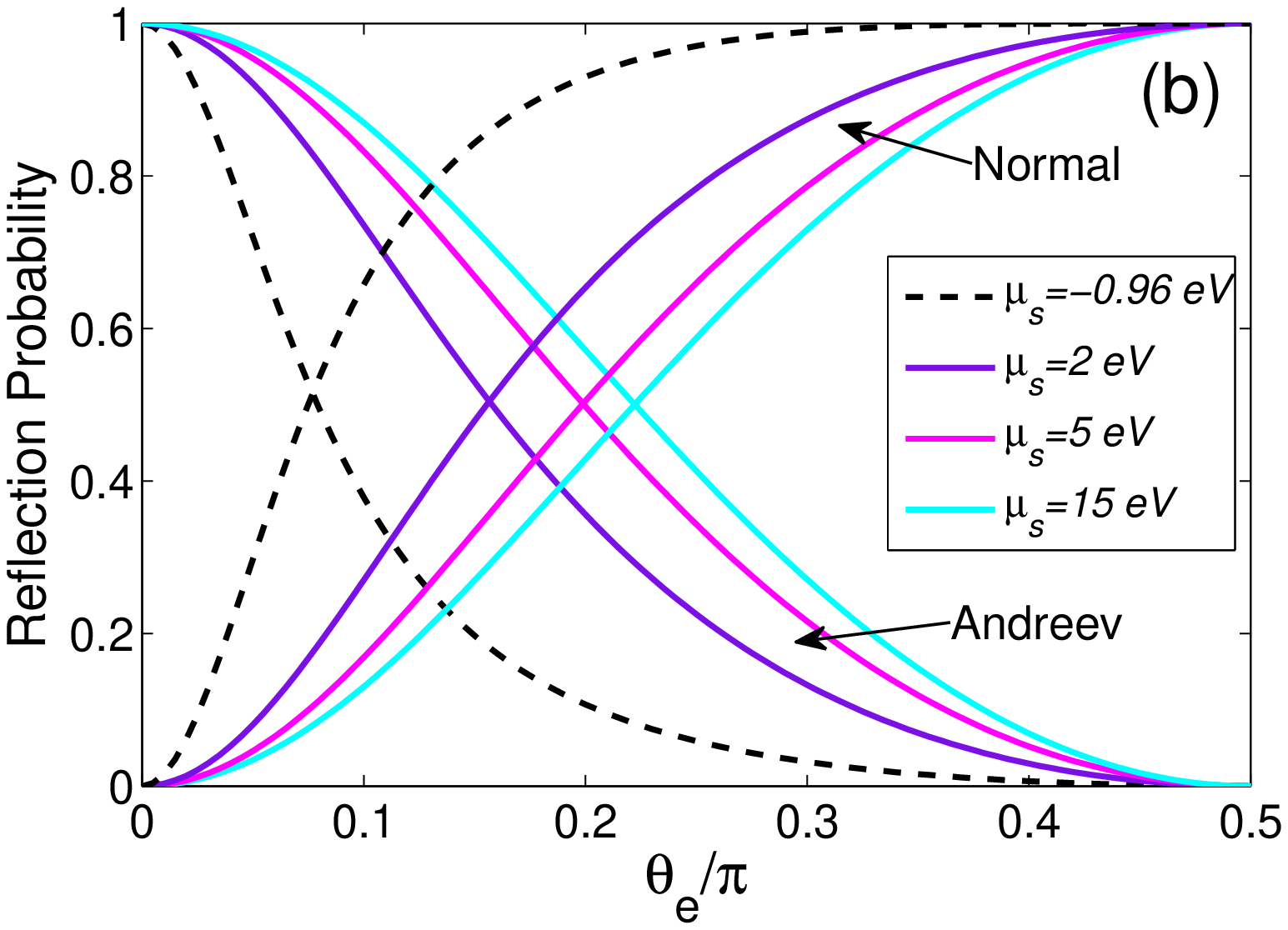}
\setcounter{figure}{1}
\caption{\footnotesize (a), (b)}
\end{center}
\end{figure}

\begin{figure}[p]
\epsfxsize=0.4 \textwidth
\begin{center}
\epsfbox{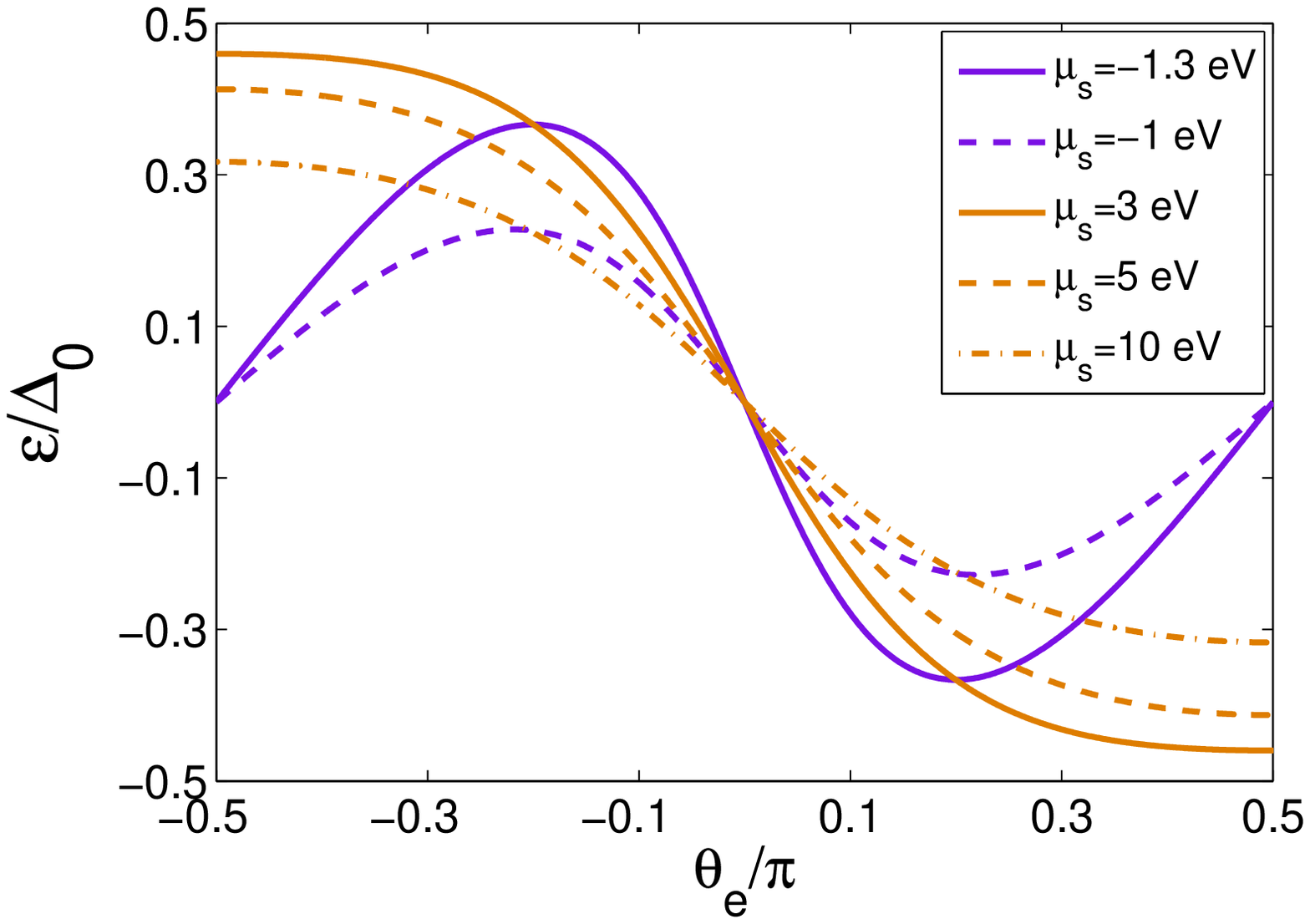}
\setcounter{figure}{2}
\caption{\footnotesize }
\end{center}
\end{figure}

\begin{figure}[p]
\epsfxsize=0.4 \textwidth
\begin{center}
\epsfbox{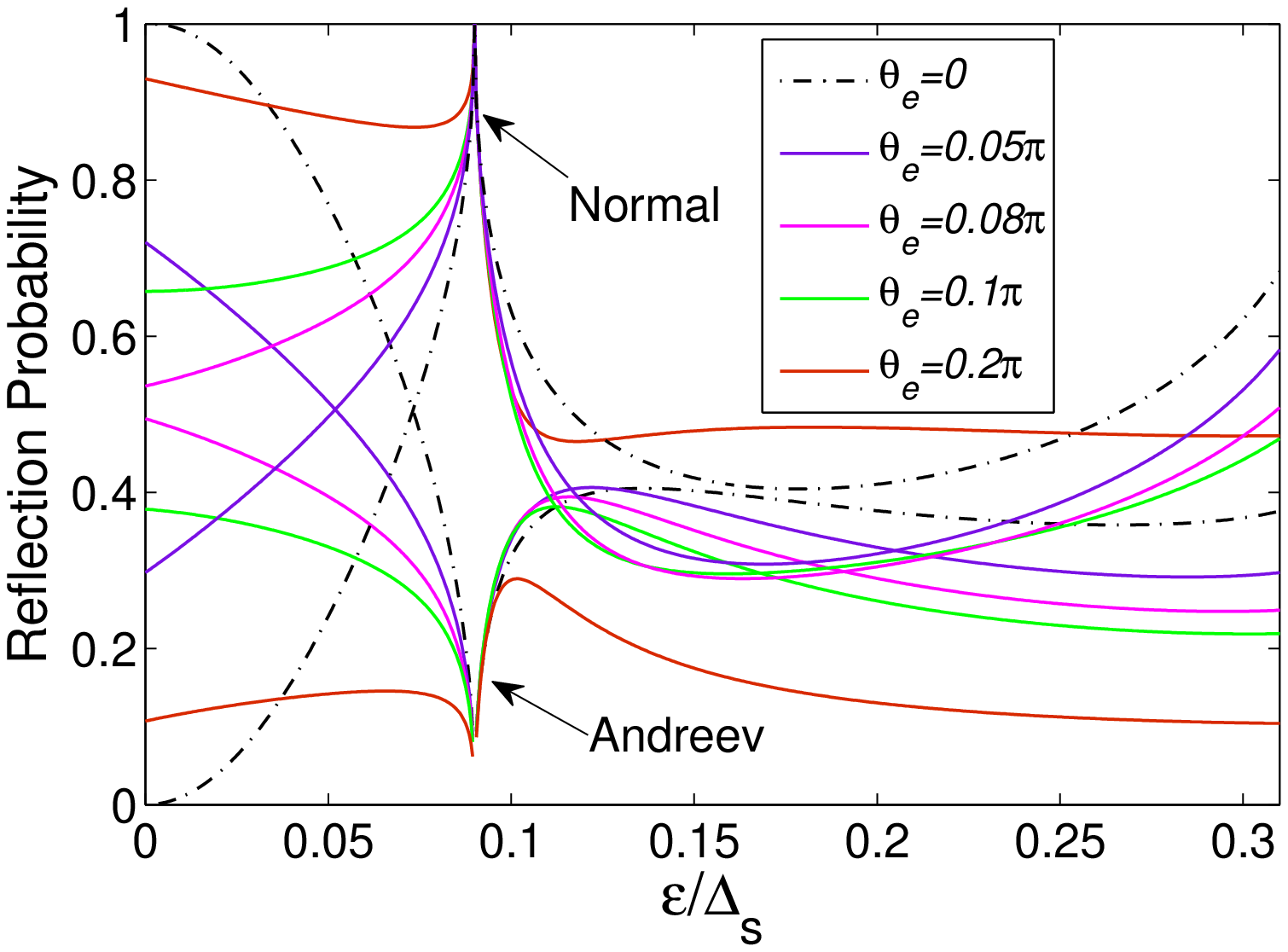}
\setcounter{figure}{3}
\caption{\footnotesize }
\end{center}
\end{figure}

\begin{figure}[p]
\epsfxsize=0.4 \textwidth
\begin{center}
\epsfbox{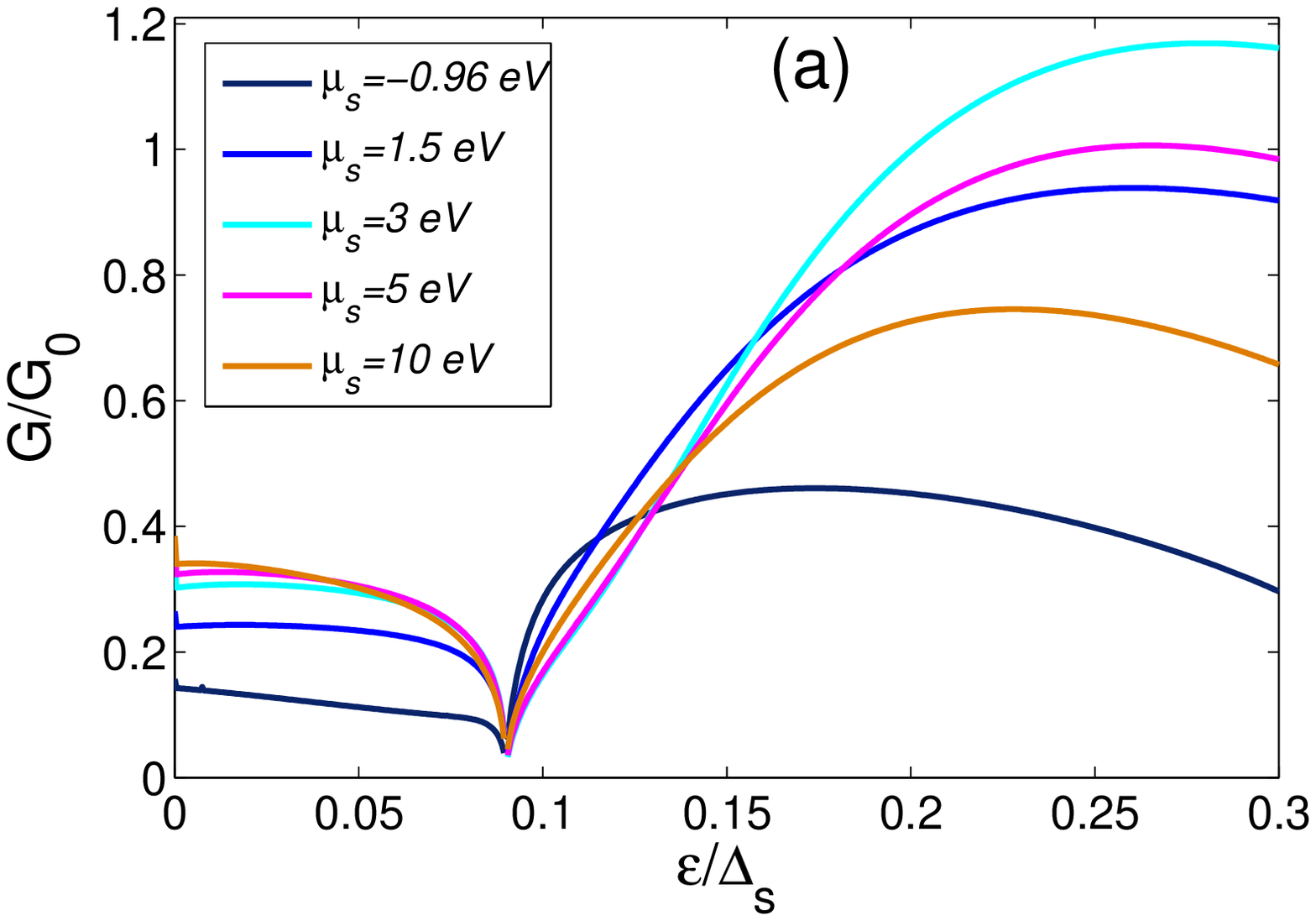}
\end{center}
\end{figure}

\begin{figure}[p]
\epsfxsize=0.4 \textwidth
\begin{center}
\epsfbox{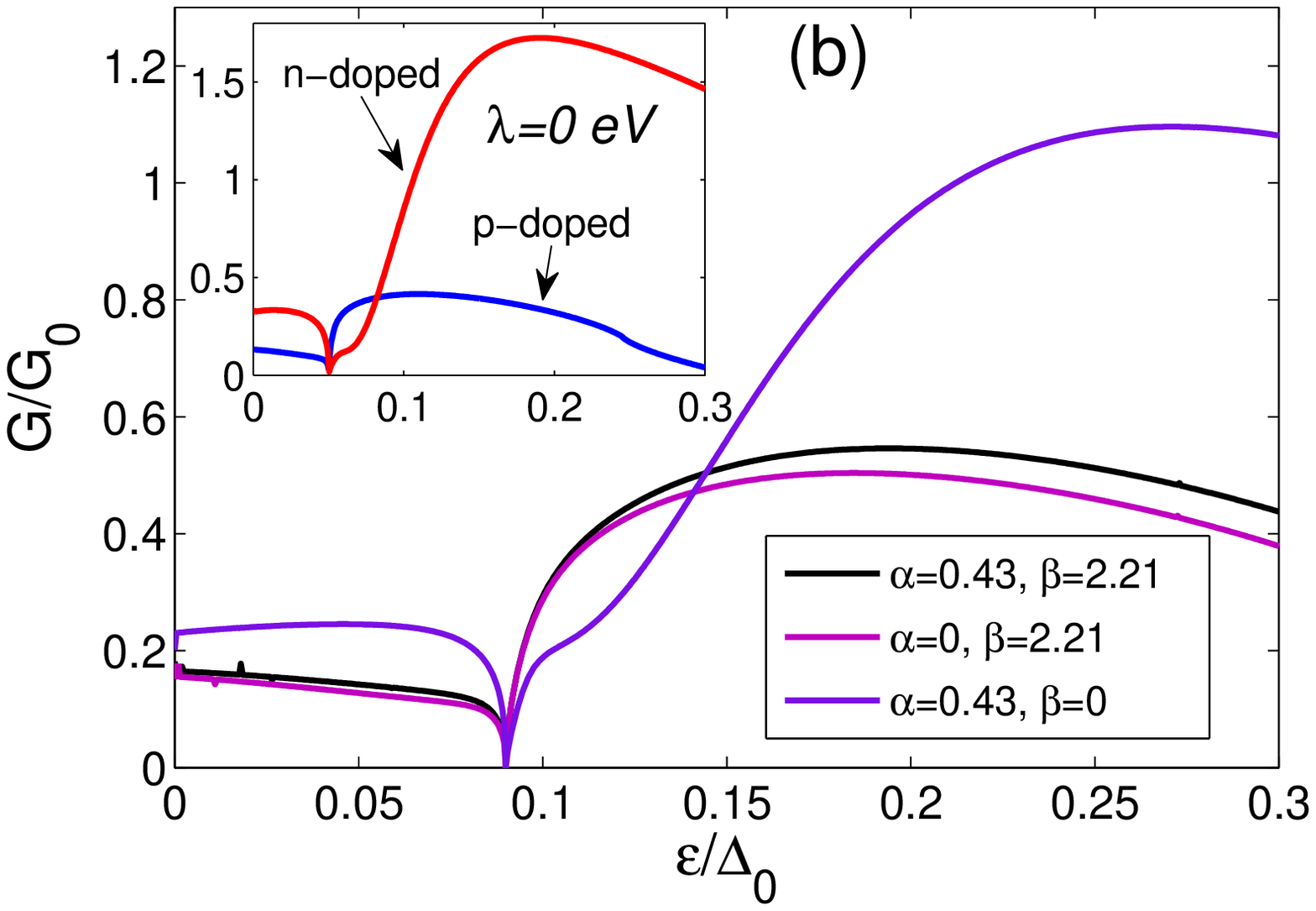}
\setcounter{figure}{4}
\caption{\footnotesize (a), (b)}
\end{center}
\end{figure}



\begin{figure}[p]
\epsfxsize=0.4 \textwidth
\begin{center}
\epsfbox{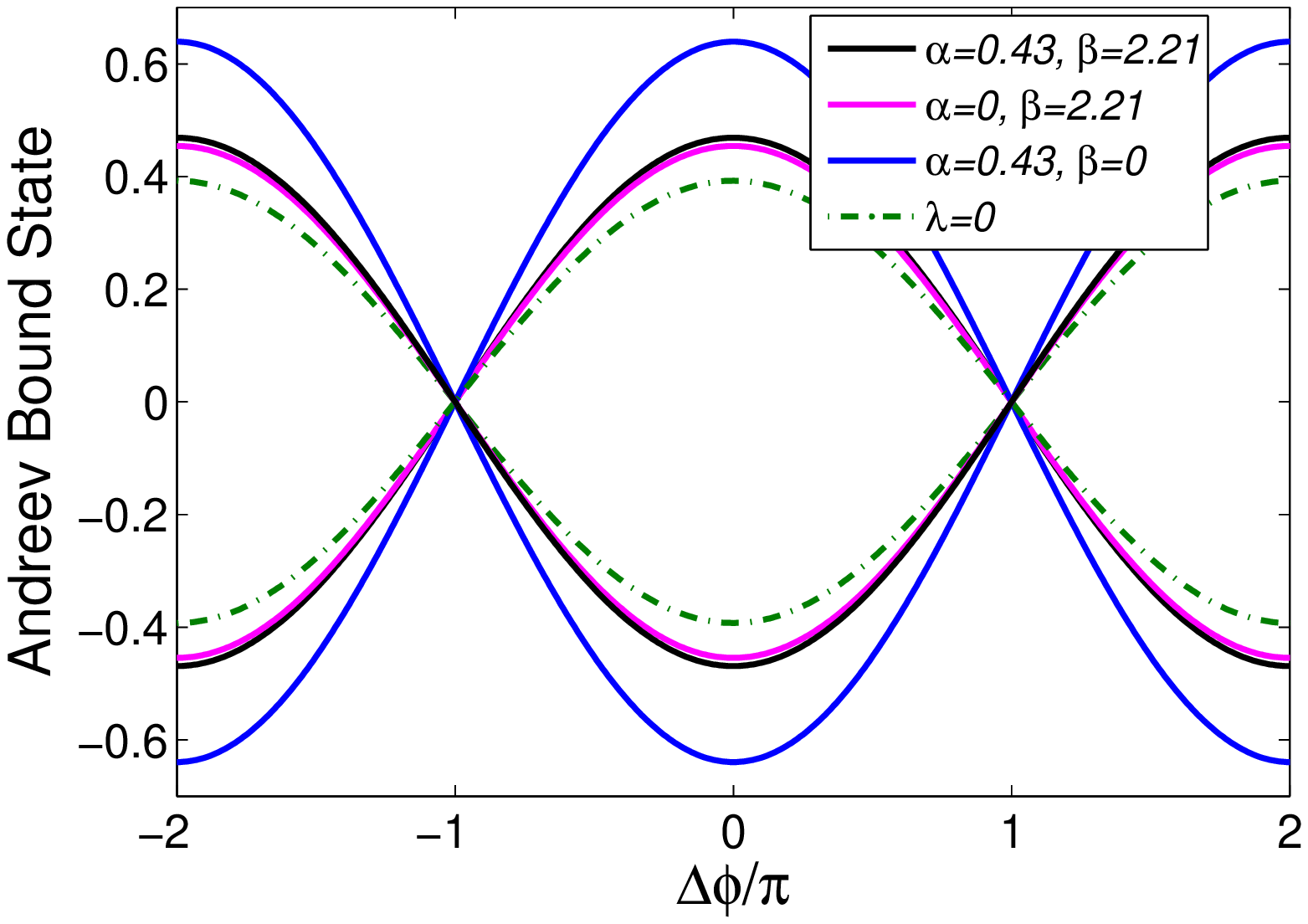}
\setcounter{figure}{5}
\caption{\footnotesize }
\end{center}
\end{figure}

\begin{figure}[p]
\epsfxsize=0.4 \textwidth
\begin{center}
\epsfbox{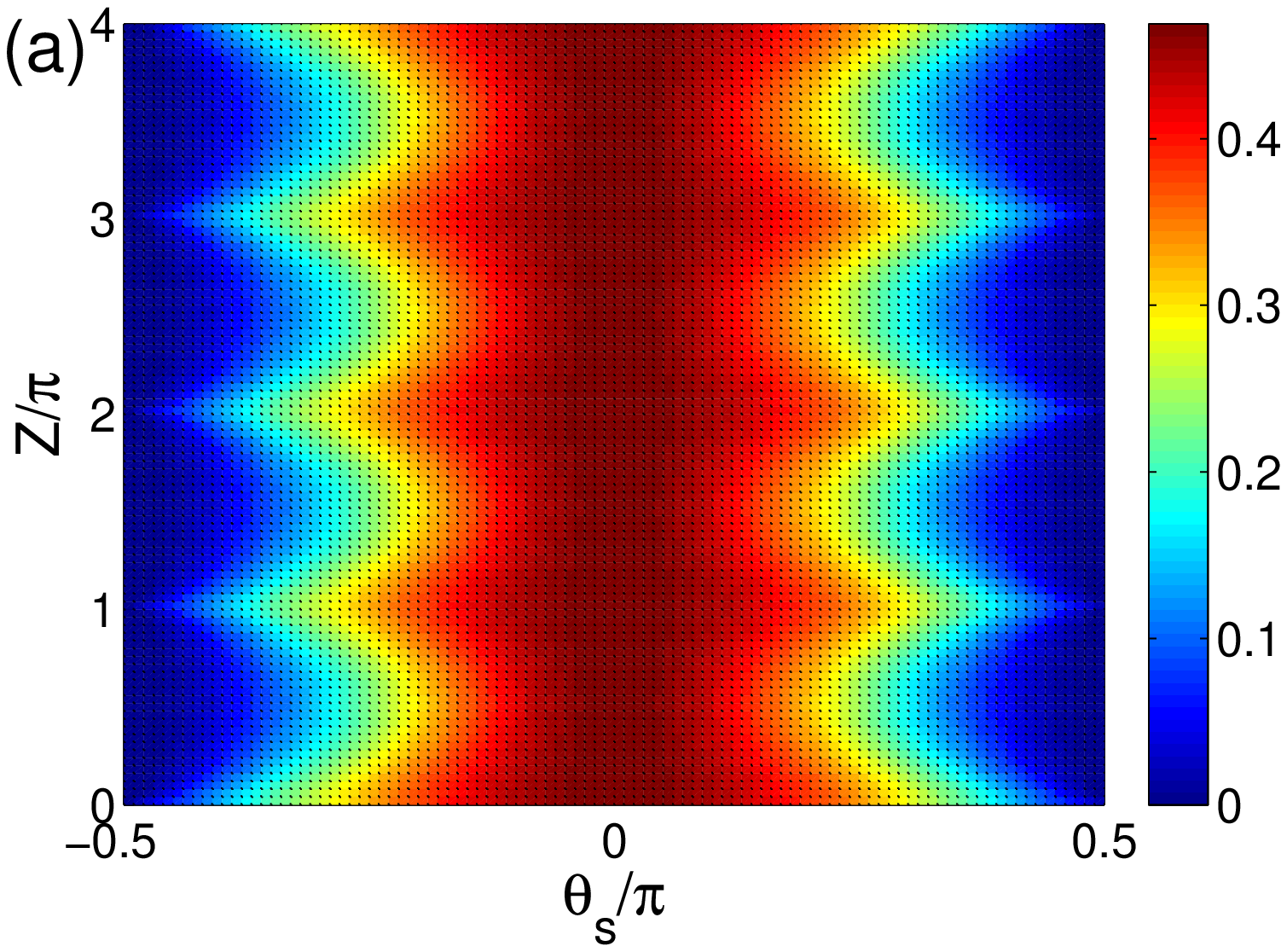}
\end{center}
\end{figure}

\begin{figure}[p]
\epsfxsize=0.4 \textwidth
\begin{center}
\epsfbox{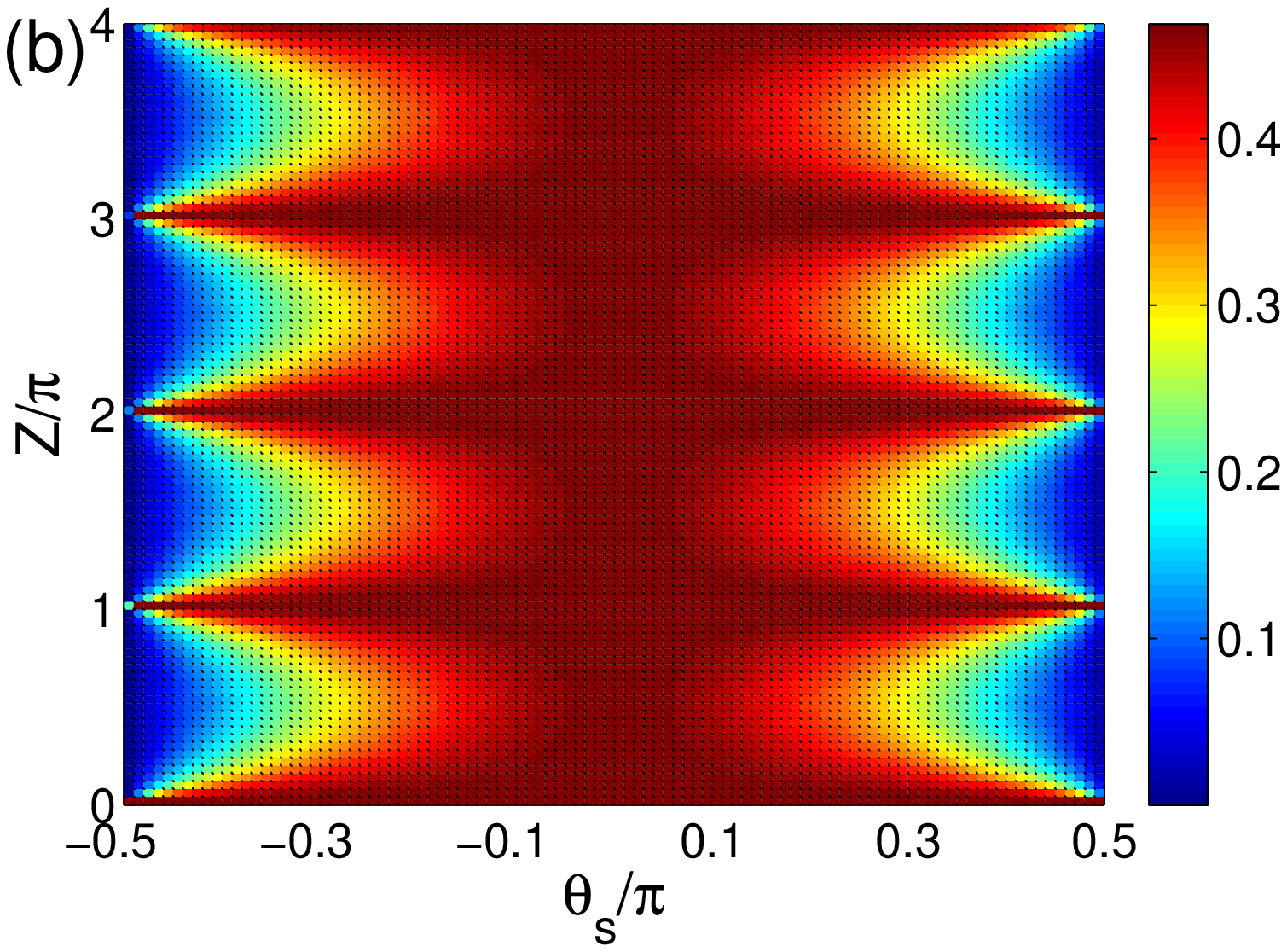}
\setcounter{figure}{6}
\caption{\footnotesize (a), (b)}
\end{center}
\end{figure}

\begin{figure}[p]
\epsfxsize=0.4 \textwidth
\begin{center}
\epsfbox{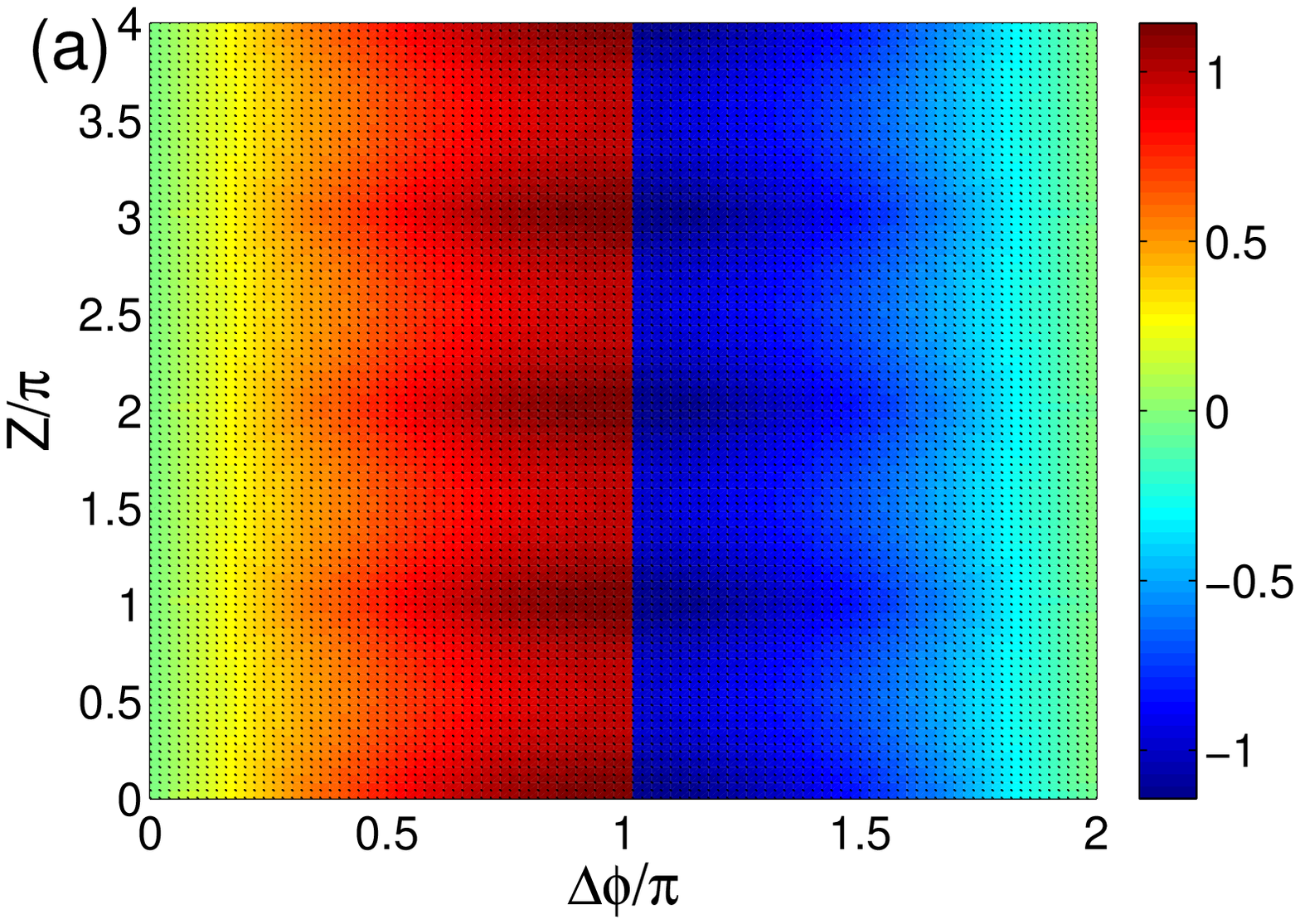}
\end{center}
\end{figure}

\begin{figure}[p]
\epsfxsize=0.4 \textwidth
\begin{center}
\epsfbox{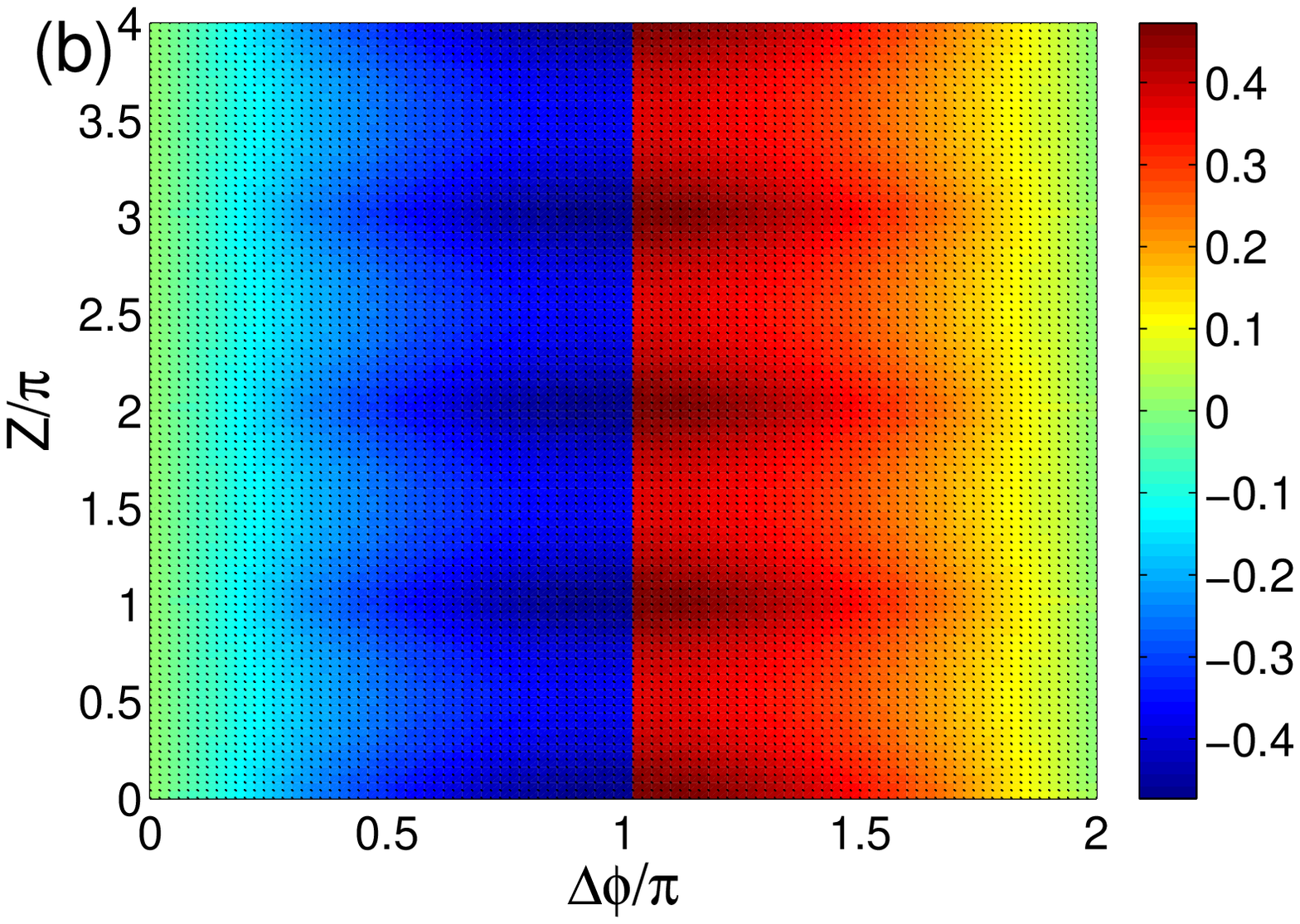}
\setcounter{figure}{7}
\caption{\footnotesize (a), (b)}
\end{center}
\end{figure}

\begin{figure}[p]
\epsfxsize=0.4 \textwidth
\begin{center}
\epsfbox{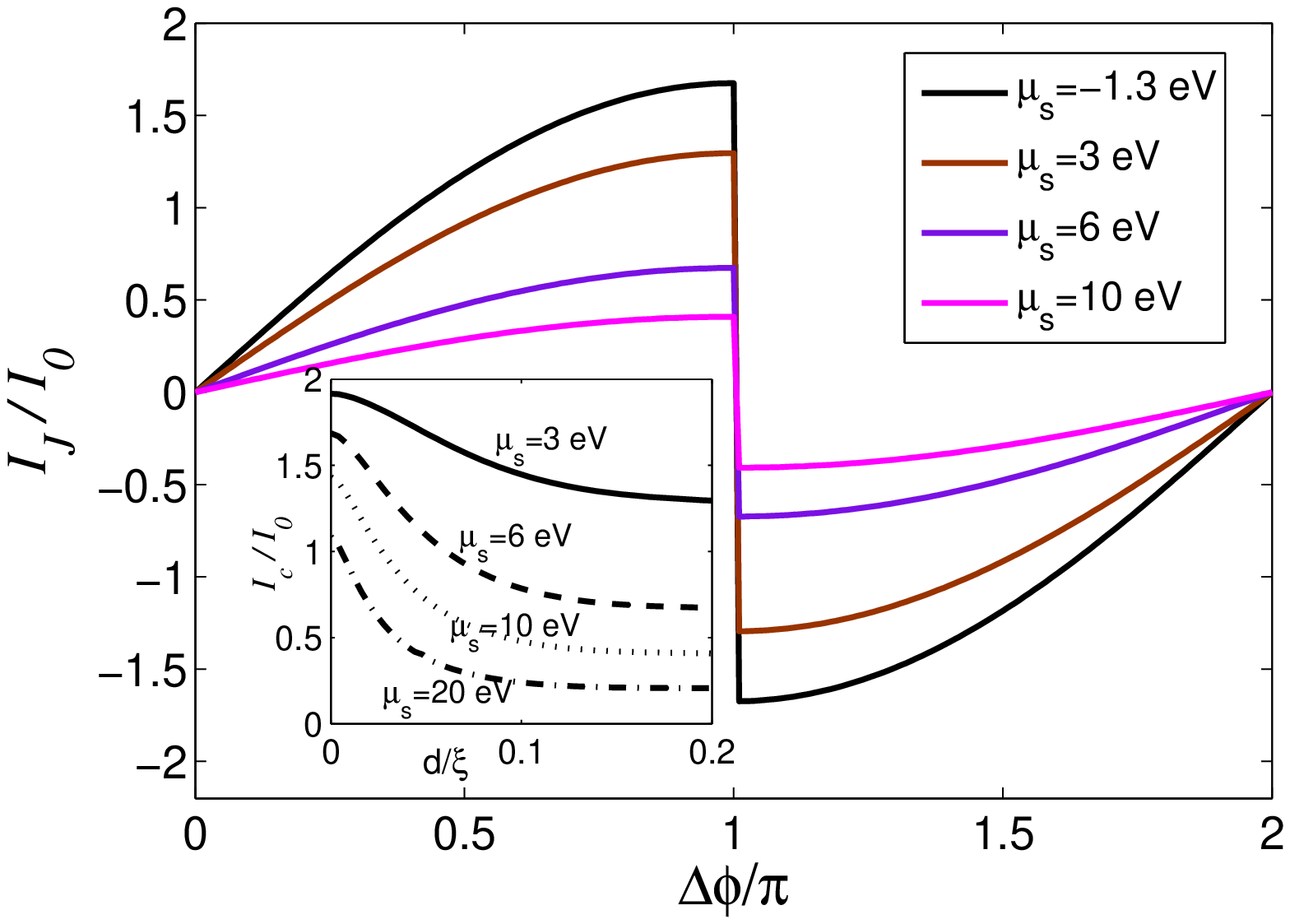}
\setcounter{figure}{8}
\caption{\footnotesize }
\end{center}
\end{figure}

\begin{figure}[p]
\epsfxsize=0.4 \textwidth
\begin{center}
\epsfbox{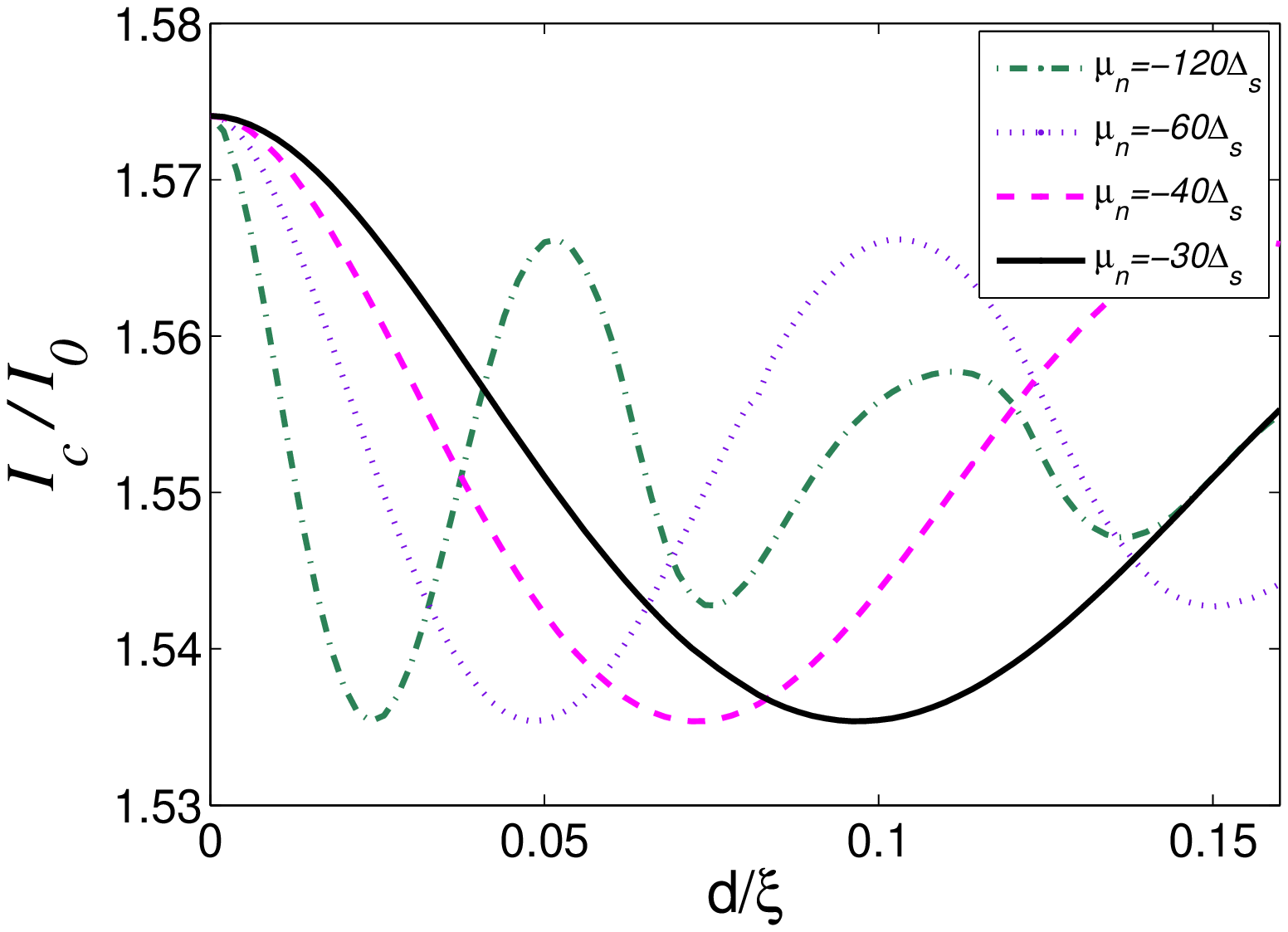}
\setcounter{figure}{9}
\caption{\footnotesize }
\end{center}
\end{figure}

\end{document}